\documentclass{emulateapj}
\usepackage{apjfonts}
\usepackage{graphicx}
\usepackage{xspace}
\usepackage{longtable}

\begin{document}

\lefthead{{\em Chandra} Observations of {\em INTEGRAL} Sources}
\righthead{Tomsick et al.}

\submitted{Accepted by ApJ}

\def\lsim{\mathrel{\lower .85ex\hbox{\rlap{$\sim$}\raise
.95ex\hbox{$<$} }}}
\def\gsim{\mathrel{\lower .80ex\hbox{\rlap{$\sim$}\raise
.90ex\hbox{$>$} }}}

\title{{\em Chandra} Observations of Eight Sources Discovered by {\em INTEGRAL}}

\author{John A. Tomsick\altaffilmark{1},
Roman Krivonos\altaffilmark{2},
Qinan Wang\altaffilmark{1},
Arash Bodaghee\altaffilmark{3},
Sylvain Chaty\altaffilmark{4,5},
Farid Rahoui\altaffilmark{6,7},
Jerome Rodriguez\altaffilmark{4},
Francesca M. Fornasini\altaffilmark{1,8}}

\altaffiltext{1}{Space Sciences Laboratory, 7 Gauss Way, University of California, 
Berkeley, CA 94720-7450, USA}

\altaffiltext{2}{Space Research Institute, Russian Academy of Sciences, Profsoyuznaya 
84/32, 117997 Moscow, Russia}

\altaffiltext{3}{Georgia College \& State University, CBX 82, Milledgeville, GA 31061, 
USA}

\altaffiltext{4}{Laboratoire AIM, UMR 7158 CEA/DSM-CNRS-Universit\'{e} Paris Diderot, 
IRFU/SAp, F-91191 Gif-sur-Yvette Cedex, France}

\altaffiltext{5}{Institut Universitaire de France, 103 Boulevard Saint-Michel, 
75005 Paris, France}

\altaffiltext{6}{European Southern Observatory, Karl Schwarzschild-Strasse
2, 85748 Garching bei Munchen, Germany}

\altaffiltext{7}{Department of Astronomy, Harvard University, 60 Garden Street, 
Cambridge, MA 02138, USA}

\altaffiltext{8}{Astronomy Department, University of California, 601 Campbell Hall, Berkeley, 
CA 94720, USA}

\begin{abstract}

We report on 0.3--10\,keV observations with the {\em Chandra X-ray Observatory} of eight hard
X-ray sources discovered within $8^{\circ}$ of the Galactic plane by the {\em INTEGRAL} satellite.
The short ($\sim$5\,ks) {\em Chandra} observations of the IGR source fields have yielded very 
likely identifications of X-ray counterparts for three of the IGR sources: IGR~J14091--6108, 
IGR~J18088--2741, and IGR~J18381--0924.  The first two have very hard spectra in the {\em Chandra} 
band that can be described by a power-law with photon indices of $\Gamma = 0.6\pm 0.4$
and --$0.7^{+0.4}_{-0.3}$, respectively (90\% confidence errors are given), and both have a unique 
near-IR counterpart consistent with the {\em Chandra} position.  IGR~J14091--6108 also displays
a strong iron line and a relatively low X-ray luminosity, and we argue that the most likely 
source type is a Cataclysmic Variable (CV), although we do not completely rule out the possibility
of a High Mass X-ray Binary.  IGR~J18088--2741 has an optical counterpart with a previously
measured 6.84\,hr periodicity, which may be the binary orbital period.  We also detect five
cycles of a possible 800--950\,s period in the {\em Chandra} light curve, which may be the
compact object spin period.  We suggest that IGR~J18088--2741 is also most likely a CV.
For IGR~J18381--0924, the spectrum is intrinsically softer with $\Gamma = 1.5^{+0.5}_{-0.4}$, 
and it is moderately absorbed, $N_{\rm H} = (4\pm 1)\times 10^{22}$\,cm$^{-2}$.  There are two 
near-IR sources consistent with the {\em Chandra} position, and they are both classified as 
galaxies, making it likely that IGR~J18381--0924 is an Active Galactic Nucleus (AGN).  For 
the other five IGR sources, we provide lists of nearby {\em Chandra} sources, which may be 
used along with further observations to identify the correct counterparts, and we discuss 
the implications of the low inferred {\em Chandra} count rates for these five sources.

\end{abstract}


\section{Introduction}

The {\em INTErnational Gamma-Ray Astrophysics Laboratory (INTEGRAL)} satellite 
\citep{winkler03} has been surveying the sky in the hard X-ray/soft gamma-ray 
band since its launch in 2002.  The Imager on-Board the {\em INTEGRAL} Satellite 
\citep[IBIS;][]{ubertini03} has detected large numbers of sources in the 20--100\,keV 
band.  The current version of the {\em INTEGRAL} General Reference 
Catalog\footnote{see http://www.isdc.unige.ch/integral/catalog/39/catalog.html} 
includes 954 sources that have been detected by IBIS, and the majority of these 
are ``IGR'' sources, meaning that either they were discovered by {\em INTEGRAL},
or they were not known to produce hard X-ray emission prior to the
{\em INTEGRAL} detection.  The ``{\em INTEGRAL} Sources'' website\footnote{see http://irfu.cea.fr/Sap/IGR-Sources/}
lists over 550 IGR sources.  The most recent published IBIS catalogs include 
\cite{bird10}, which includes information on more than 700 sources (IGR and non-IGR), 
and \cite{krivonos12}, which lists 402 sources within $17.5^{\circ}$ of the Galactic 
plane\footnote{see http://hea.iki.rssi.ru/integral}.

Thus, the IGR sources represent a large population of proven hard X-ray emitters, but, 
in most cases, more information is needed to determine their nature.  The fact that
these sources produce emission above 20\,keV indicates that they are sites of particle 
acceleration or extreme heating, which leads to the production of non-thermal emission.  
Based on the source type identifications that have been obtained to date, Active Galactic 
Nuclei (AGN) are the most numerous group of IGR sources, but there are hundreds of Galactic 
sources as well, including Low Mass X-ray Binaries (LMXBs), High Mass X-ray Binaries 
(HMXBs), Cataclysmic Variables (CVs), and Pulsar Wind Nebulae (PWNe).  {\em INTEGRAL} 
discoveries have increased the total number of known HMXBs from 65 to 96 and have
multiplied the number of supergiant HMXBs almost threefold from 13 to 36 \citep{walter15}.  
Many of the IGR HMXBs belong to two new classes: the obscured HMXBs, where the compact 
object is enshrouded in the wind from the companion star \citep[e.g.,][]{mg03,fc04,walter06}; 
and the Supergiant Fast X-ray Transients 
\citep[SFXTs;][]{negueruela06,smith06,pellizza06,sidoli13,romano14}.  
There are many other interesting individual Galactic objects such as the LMXB and transitional 
pulsar IGR~J18245--2452 \citep{papitto13}, and the high velocity pulsar and PWN IGR~J11014--6103 
\citep{pavan11,tomsick12}.

Observations with the {\em Chandra X-ray Observatory} are especially useful for
determining the nature of the IGR sources because {\em Chandra}'s superior angular 
resolution can improve the source localization from arcminutes (with {\em INTEGRAL})
to less than an arcsecond; thus, providing the opportunity to identify counterparts at
other wavelengths, such as the optical or near-IR where all-sky images and catalogs
are readily available.  In addition, {\em Chandra}'s 0.3--10\,keV coverage provides
constraints on the soft X-ray spectrum, allowing for a determination of column 
densities, spectral slopes, and, in some cases, emission lines.  We have been 
carrying out {\em Chandra} programs to follow up IGR sources in several previous 
{\em Chandra} observing cycles
\citep{tomsick06,tomsick08,tomsick09a,tomsick12a,bodaghee12a}, and other groups also
have similar programs \citep[e.g.,][]{fiocchi10,ratti10,paizis11,paizis12,nowak12,karasev12}.

For the {\em Chandra} cycle 15 program, we selected IGR sources from the 
\cite{krivonos12} catalog.  We excluded sources whose classification (e.g., AGN, HMXB, 
etc.) is known, leaving 34 unclassified sources of which 30 have an IGR designation.  
Next, we only considered sources situated within $8^{\circ}$ of the Galactic Plane, which 
increases the probability of identifying new X-ray binaries, CVs, and PWNe.  This 
selection also takes advantage of {\em Chandra}'s unparalleled X-ray positional accuracy, 
which we require to pinpoint an optical/IR counterpart in these crowded fields.  After
removing sources with existing soft X-ray coverage with {\em Chandra} and {\em XMM-Newton}, 
we were left with ten sources.  We reported on the results of the cycle 15 {\em Chandra} 
observations for two of the sources (IGR~J04059+5416 and IGR~J08297--4250), concluding 
that they are AGN \citep{tomsick15}

In this paper, we report on the results of the remaining eight observations.  Section 2
provides a brief description of the {\em Chandra} observations and how the data were
processed.  In Section 3, we present results on {\em Chandra} source detection and 
photometry.  For the three {\em Chandra} sources that we identify with IGR sources, 
we present {\em Chandra} and {\em INTEGRAL} energy spectra, {\em Chandra} light
curves, and IR source identifications.  A discussion of the nature of the three sources 
is included in Section 4, and conclusions are presented in Section 5.  In addition, we 
include an Appendix with a full listing of the {\em Chandra} sources detected.

\section{Chandra Observations and Data Processing}

Information about the eight {\em Chandra} observations, which occurred between late-2013 
and early-2015, is given in Table~\ref{tab:obs}.  The integration time for each observation 
is $\sim$5\,ks, and we used the Advanced CCD Imaging Spectrometer (ACIS-I) instrument 
\citep{garmire03}.  The pointing position was at the best known source position, and, in 
seven cases, this corresponds to the position measured by {\em INTEGRAL} \citep{krivonos12}.  
For IGR~J14091--6108, a likely {\em Swift} counterpart has been reported \citep{landi12}, 
and we used the {\em Swift} position for the {\em Chandra} pointing.  After obtaining the 
data from the {\em Chandra} X-ray Center, all of the {\em Chandra} data reduction in this 
work was done using the {\em Chandra} Interactive Analysis of Observations (CIAO) 
version 4.7 software and Calibration Data Base (CALDB) 4.6.7.  We made new event lists 
using {\ttfamily chandra\_repro} and used these event lists for the analysis described 
below.

\section{Analysis and Results}

\subsection{Chandra Source Detection and Photometry}

For each of the eight observations, we inspected the full field of view (ACIS-I 
and ACIS-S) and used {\ttfamily wavdetect} \citep{freeman02} to search for 
{\em Chandra} sources over ACIS-I, on which the {\em INTEGRAL} source position
falls in each case.  The accuracy of the {\em INTEGRAL} positions is 
$2.\!^{\prime}1$ at 68\% confidence \citep{krivonos12}, indicating that it is
very likely that the true source position lies in the 
$16.\!^{\prime}9$-by-$16.\!^{\prime}9$ ACIS-I field of view.  In order to detect 
sources in both soft and hard bands, we divided events into a soft band with 
energy between 0.3\,keV and 2\,keV, and a hard band with energy between 2\,keV 
and 10\,keV. Before we ran {\ttfamily wavdetect}, we used {\ttfamily fluximage} 
and {\ttfamily mkpsfmap} to generate the exposure map and the point spread function 
(PSF) map in order to improve the accuracy of detection.  A small number of 
sources with one or two counts was detected, and we discarded these.  In total, 
we detected 115 sources in the ACIS-I fields of view, with the most and least 
crowded fields having 29 and 7 sources, respectively (see Table~\ref{tab:obs} 
for the totals in each field).  Tables listing the basic information (position, 
counts detected, and hardness) for all 115 sources, as well as more details on
how we obtained this information, are provided in the Appendix.

\begin{figure}
\includegraphics[scale=0.5]{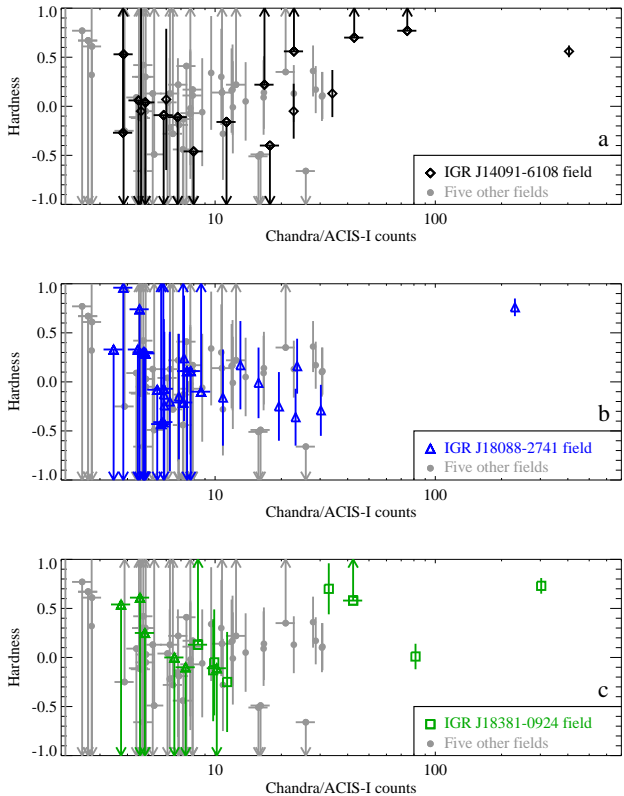}
\caption{\footnotesize Hardness-intensity diagrams, where the intensity is given as the number 
of ACIS-I counts in the 0.3--10\,keV energy band.  The hardness is given by $(C_{2}-C_{1})/(C_{2}+C_{1})$, 
where $C_{2}$ is the number of counts in the 2--10\,keV band and $C_{1}$ is the number of counts 
in the 0.3--2\,keV band.  The black diamonds in {\em (a)}, blue triangles in {\em (b)}, and green 
squares in {\em (c)} correspond to sources in the IGR~J14091--6108, IGR~J18088--2741, and 
IGR~J18381--0924 fields, respectively.  The points in the other fields are shown in grey.\label{fig:hi}}
\end{figure}

For all 115 sources, we performed photometry to determine the number of source
counts in the 0.3--10\,keV, 0.3--2\,keV, and 2--10\,keV bands.  The size of 
the {\em Chandra} PSF changes significantly over the field of view, and we 
determined the off-axis angle for each source and then used the {\em Chandra} 
PSF Viewer\footnote{see http://cxc.cfa.harvard.edu/cgi-bin/prop\_viewer/build\_viewer.cgi?psf}
to determine the radius of the 90\% encircled counts fraction at 3\,keV for
each source.  For the photometry, we extracted counts from circular regions
that were twice this radius, so that the extraction regions enclosed more 
than 99\% of the counts.  For each field, we identified a large source-free
region, and used this region to estimate and then subtract the background
contribution within each source extraction region.
The results of the photometry are shown by plotting the hardness of each
source, which is defined as $(C_{2}-C_{1})/(C_{2}+C_{1})$, where $C_{2}$ is
the number of counts in the 2--10\,keV band and $C_{1}$ is the number of 
counts in the 0.3--2\,keV band, vs. the intensity, which is the number of
counts in the 0.3--10\,keV band (Figure~\ref{fig:hi}).

Three sources stand out from the rest of the sources in the hardness-intensity 
diagram:  CXOU~J140846.0--610754 in the field of IGR~J14091--6108; 
CXOU~J180839.8--274131 in the field of IGR~J18088--2741; and CXOU~J183818.5--092552 
in the field of IGR~J18381--0924.  CXOU~J140846.0--610754 has 404.9 counts, a 
hardness of $0.56\pm 0.06$, and it is only $0.\!^{\prime}45$ from the center 
of the error circle, which is well within the 1-$\sigma$ {\em INTEGRAL} error 
circle.  CXOU~J180839.8--274131 has 229.9 counts and a hardness of $0.76\pm 0.09$.  
While it is $4.\!^{\prime}5$ from the center of the {\em INTEGRAL} error circle, it 
is an order of magnitude brighter than any source that is closer to the {\em INTEGRAL} 
position, and it is harder than any other source in the field with more than 10 counts.  
CXOU~J183818.5--092552 has 302.7 counts, a hardness of $0.73\pm 0.08$, and it is 
$2.\!^{\prime}8$ from the best {\em INTEGRAL} position, which is within the 
2-$\sigma$ error circle.  Thus, these three {\em Chandra} sources are very likely 
counterparts to their respective {\em INTEGRAL} sources, and in the following 
sections, we report on the details of these three sources, including their 
{\em Chandra} and {\em INTEGRAL} energy spectra, their {\em Chandra} light curves, 
and whether their {\em Chandra} positions allow us to identify counterparts at
other wavelengths.

For the other five IGR fields, some of the {\em Chandra} sources detected are
potential counterparts.  However, in those cases, there is a significant probability
that the association is spurious.  We first focus on CXOU~J140846.0--610754, 
CXOU~J180839.8--274131, and CXOU~J183818.5--092552 and then consider the upper 
limits for the other five fields.

\subsection{Chandra and INTEGRAL Energy Spectra}

For all three sources, we produced {\em Chandra} and {\em INTEGRAL} energy spectra.  
For {\em Chandra}, we used circular source extraction regions with the same
radii used for the photometry: $2.\!^{\prime\prime}5$ for CXOU~J140846.0--610754 and 
CXOU~J183818.5--092552; and $5.\!^{\prime\prime}0$ for CXOU~J180839.8--274131. 
However, as CXOU~J140846.0--610754 was on-axis with an ACIS count rate of 
0.08\,c/s, the spectrum can be distorted by photon pile-up.  Using the 
Portable, Interactive Multi-Mission Simulator 
(PIMMS)\footnote{see http://asc.harvard.edu/toolkit/pimms.jsp}, we estimate
the pile-up level at 10\%.  Thus, for CXOU~J140846.0--610754, we used an annular
source region with an inner radius of $0.\!^{\prime\prime}5$ and an outer radius 
of $2.\!^{\prime\prime}5$. In all three cases, a background spectrum was obtained 
from a source-free region close to the source.  The source and background spectra 
and the response matrices were produced with the {\ttfamily specextract} script.  
We included the 0.3--10\,keV energy range, and rebinned each spectrum with the 
criterion that the source be detected at a signal-to-noise (S/N) level of 3 or 
greater in each bin (except for the highest energy bin).

The {\em INTEGRAL} energy spectra were produced using publicly available data from the 
IBIS/ISGRI instrument over a time period from 2003 to the end of 2014, yielding
effective exposure times\footnote{These correspond to dead-time corrected exposure
times calculated for the fully coded field of view.} of 5.0, 23.6, and 8.0\,Ms for 
IGR~J14091--6108, IGR~J18088--2741, and IGR~J18381--0924, respectively.  
We reduced the IBIS/ISGRI data with the 
{\em INTEGRAL} data analysis package developed at IKI RAN\footnote{Space Research 
Institute of the Russian Academy of Sciences, Moscow, Russia} 
\citep[see, e.g.,][and references therein]{churazov05,krivonos10,churazov14} using the
most up-to-date gain calibration for ISGRI \citep{caballero13} available through the 
Offline Scientific Analysis (OSA) version 10.1, provided by the 
ISDC\footnote{INTEGRAL Data Center for Astrophysics, http://www.isdc.unige.ch/}. 
To take the ongoing detector degradation and loss of sensitivity at low energies
\citep{caballero13} into account, we adjusted the flux scale in each IBIS sky image
using the flux of the Crab nebula measured in the observation that is closest in time
to when the data for the sky image was obtained.  This results in a smooth recalibration 
of the ancillary response function over the span of the observations from 2003 to 2014.
We previously used the same procedure for a subset of the data as described in 
\cite{krivonos12}.  The energy spectra were obtained from sky mosaic images in four 
energy bands: 17--26\,keV, 26--38\,keV, 38--57\,keV, and 57--86\,keV. The corresponding 
energy response matrices were produced using Crab nebula observations, assuming a 
spectral shape of $10 (E/1 {\rm keV})^{-2.1}$\,photons\,cm$^{-2}$\,s$^{-1}$\,keV$^{-1}$.

\begin{figure*}
\begin{center}
\includegraphics[scale=0.9]{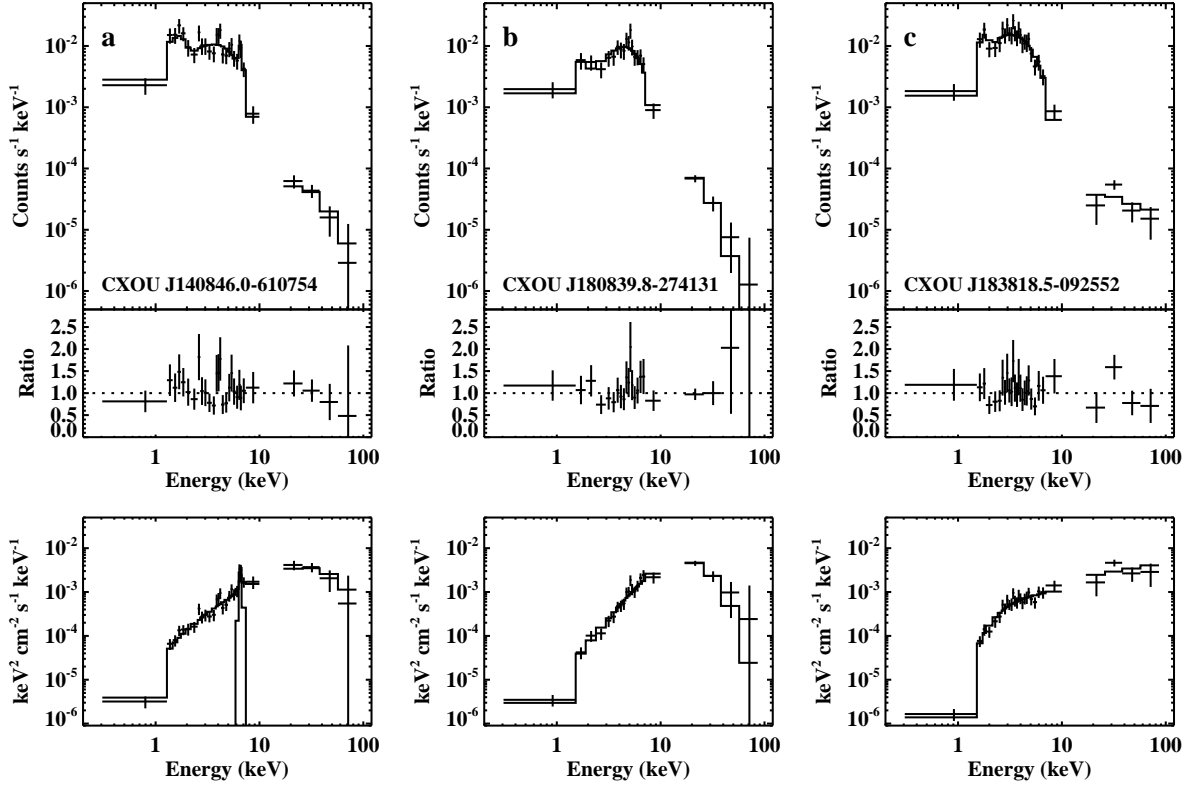}
\end{center}
\caption{{\em Chandra} and {\em INTEGRAL} energy spectra for ({\em a}) CXOU~J140846.0--610754/IGR~J14091--6108, 
({\em b}) CXOU~J180839.8--274131/IGR~J18088--2741, and ({\em c}) CXOU~J183818.5--092552/IGR~J18381--0924.  For
each source, the top panel is the counts ``folded'' spectrum, the middle panel is the data-to-model ratio, and
the bottom panel is the ``unfolded'' spectrum in flux units.  From left to right, the models are: 
{\ttfamily tbabs*(gaussian+cutoffpl)}, {\ttfamily tbabs*cutoffpl}, and {\ttfamily tbabs*powerlaw}.
\label{fig:spectra}}
\end{figure*}

For each source, we initially fit the {\em Chandra} and {\em INTEGRAL} spectra
separately.  The S/N = 3 binning for {\em Chandra} results in about 10 counts per bin, 
which is close to the Poisson regime.  Thus, for the {\em Chandra}-only fits, the
fitting was done by minimizing the Cash statistic \citep{cash79}.  Due to the high
{\em INTEGRAL} background, we performed the fits to the {\em INTEGRAL} spectra and 
the fits to the joint {\em Chandra}+{\em INTEGRAL} spectra by minimizing the 
$\chi^{2}$ statistic.  The fit parameters and $\chi^{2}$ values are reported in 
Table~\ref{tab:parameters}, and the folded and unfolded spectra are shown in 
Figure~\ref{fig:spectra}.  The errors on the spectral parameters in the table as 
well as values quoted in this section are at the 90\% confidence level.

For CXOU~J140846.0--610754, an absorbed power-law model to the {\em Chandra} spectrum
leaves residuals in the iron line region, and we find that the addition of a Gaussian
iron line at $E_{\rm line} = 6.6\pm 0.2$\,keV with a width of $\sigma_{\rm line} < 0.6$\,keV 
leads to an improvement in the Cash statistic from 32.6 for 22 degrees of freedom (dof) 
to 19.4 for 19 dof.  We used {\tt simftest} to produce 10,000 simulated {\em Chandra} 
spectra with the absorbed power-law model, fit the spectra with and without the addition 
of a Gaussian with $E_{\rm line}$ between 6.4 and 7.1\,keV and $\sigma_{\rm line} < 0.6$\,keV, 
and recorded the Cash statistics for all the fits.  Out of the 10,000 spectra, an 
improvement in the Cash statistic of 13.2 (=32.6--19.4) or more was seen eight times, 
corresponding to a detection significance of 3.35-$\sigma$.  Thus, we included this line, 
which has an equivalent width (EW) of $1.2^{+0.9}_{-0.7}$\,keV, in our subsequent fits.  
Some weak residuals remain near 4.0\,keV, but adding a Gaussian leads to only a small
reduction in the Cash statistic to 15.0 for 16 dof.  Using {\tt simftest}, there is
a 15\% probability of obtaining such an improvement by chance, indicating that the
4\,keV feature is not statistically significant.

As reported in Table~\ref{tab:parameters}, with {\em Chandra}, we measure a column density 
of $(1.0^{+0.7}_{-0.6})\times 10^{22}$\,cm$^{-2}$ and a power-law photon index of $\Gamma = 0.6\pm 0.4$ 
for CXOU~J140846.0--610754.  This indicates that the spectrum in the 0.3--10\,keV bandpass
is even harder than the previously reported {\em Swift} value of $\Gamma\sim 1.3$ \citep{landi12}.
While the photon index measured by {\em Chandra} is very hard, 
a power-law fit to the {\em INTEGRAL} spectrum yields a much softer photon index of 
$\Gamma = 2.9^{+1.0}_{-0.7}$, indicating that there must be a break or a cutoff between the
{\em Chandra} and {\em INTEGRAL} spectra.  To fit the {\em Chandra} and {\em INTEGRAL} spectra 
together, we used the model {\tt cutoffpl}, which is a power-law multiplied by an exponential, 
$e^{-E/E_{\rm fold}}$.  As the {\em Chandra} and {\em INTEGRAL} observations were made at different 
times, and the source might be variable, we originally allowed for a normalization difference 
between the two spectra, obtaining $N_{INTEGRAL}/N_{Chandra} = 1.8^{+2.8}_{-1.1}$.  
Table~\ref{tab:parameters} shows that the changes in the parameters are not significant when we 
fix the normalization ratio to 1.0.  With the ratio fixed to 1.0, we find $\Gamma = 0.1^{+0.4}_{-0.5}$, 
and $E_{\rm fold} = 14^{+11}_{-5}$\,keV, and the folded and unfolded spectra are shown in 
Figure~\ref{fig:spectra}.  For these fits, we restricted $\sigma_{\rm line}$ to be less than
0.6\,keV.  

We performed similar spectral fits for CXOU~J180839.8--274131, and the results are reported
in Table~\ref{tab:parameters}.  While the {\em Chandra} power-law index is
$\Gamma$ = --$0.7^{+0.4}_{-0.3}$, fitting the {\em INTEGRAL} spectrum separately gives 
$\Gamma = 3.8^{+1.2}_{-0.8}$.  As the {\em Chandra} source is $4.8^{\prime}$ off-axis, and 
the ACIS count rate is lower than for CXOU~J140846.0--610754, 0.05\,c/s, photon pile-up
is not a concern, indicating that the spectrum is very hard in the 0.3--10\,keV band
and has a strong cutoff at higher energies.  Fitting the {\em Chandra} and {\em INTEGRAL}
spectra with a {\tt cutoffpl} model and the normalization ratio free gives 
$N_{INTEGRAL}/N_{Chandra} = 0.8^{+0.6}_{-0.4}$.  With the normalization fixed to 1.0, we find 
$\Gamma$ = --$1.5^{+0.4}_{-0.2}$ and $E_{\rm fold} = 4.8^{+1.1}_{-0.8}$\,keV.  We also find that 
the spectrum is not highly absorbed with $N_{\rm H} < 7\times 10^{21}$\,cm$^{-2}$.  

The CXOU~J183818.5--092552 spectrum is different from the previous two sources in that
the power-law slopes measured by {\em Chandra} ($\Gamma = 1.5^{+0.5}_{-0.4}$) and
{\em INTEGRAL} ($\Gamma = 1.8\pm 0.6$) are consistent with each other.  
Fitting the {\em Chandra} and {\em INTEGRAL} spectra together gives a normalization 
ratio of $N_{INTEGRAL}/N_{Chandra} = 1.6^{+2.1}_{-0.9}$.  With the normalization ratio fixed 
to 1.0, we find $\Gamma = 1.4\pm 0.1$, and no cutoff is required (see Figure~\ref{fig:spectra}).  
This source also differs from the other two in that it is moderately absorbed with 
$N_{\rm H} = (3.8^{+0.9}_{-0.7})\times 10^{22}$\,cm$^{-2}$.

\subsection{Chandra Light Curves}

\begin{figure*}
\begin{center}
\includegraphics[scale=0.9]{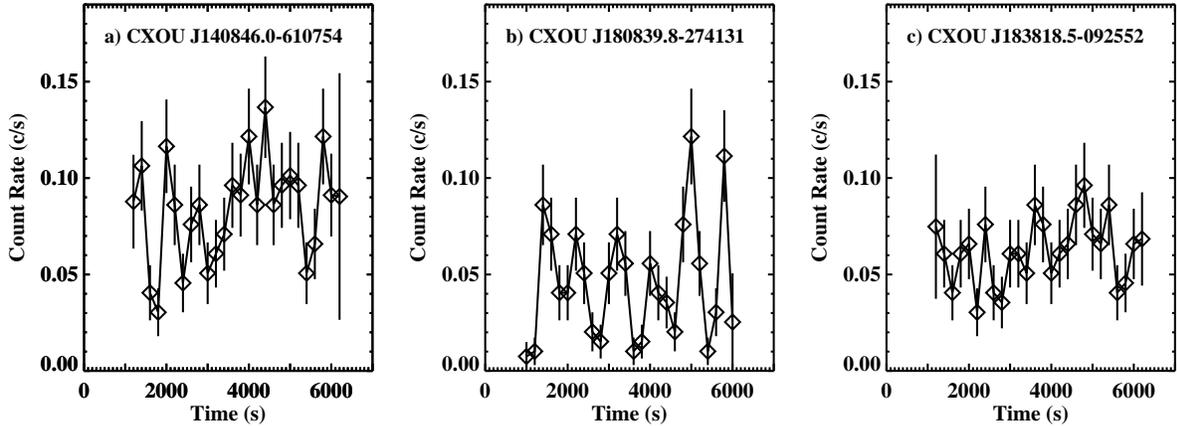}
\end{center}
\caption{{\em Chandra} 0.3--10\,keV light curves with 200\,s time bins for ({\em a}) CXOU~J140846.0--610754, 
({\em b}) CXOU~J180839.8--274131, and ({\em c}) CXOU~J183818.5--092552.  \label{fig:lc_chandra}}
\end{figure*}

For the three {\em Chandra} sources associated with {\em INTEGRAL} sources, we made 
0.3--10\,keV light curves with 200\,s time bins (Figure~\ref{fig:lc_chandra}).  We performed 
two tests to check the sources for variability: a Kolmogorov-Smirnov (KS) test and a $\chi^{2}$ 
test.  The KS test is good for low count rates because it compares the event arrival times to 
a constant reference distribution and does not require the data to be binned.  While the KS 
test is sensitive to random variations, like bursts, it is not as good at finding periodic 
variability.  Thus, we also perform a $\chi^{2}$ test by fitting a constant function to the 
binned light curves.  Although this test is problematic in low count situations where the time 
bins follow a Poisson rather than a Gaussian distribution, in our case, we typically have
10--20 counts per bin, and $\chi^{2}$ statistics are a good approximation.  

For CXOU~J140846.0--610754, the KS test measures only a 1.2\% probability that the source
is constant.  The $\chi^{2}$ test also indicates that the source is very likely to be
variable as the reduced $\chi^{2}$ is 2.1 for 25 dof.  Although the light curve for
CXOU~J180839.8--274131 appears to be even more variable, the KS test probability is higher, 
2.2\%.  However, the fit to the binned light curve with a constant gives a reduced
$\chi^{2}$ of 4.1 for 25 dof, confirming that there is a high level of variability.
For the third source, CXOU~J183818.5--092552, the KS test probability is 8.5\%, which 
does not indicate significant variability.  This is confirmed by the second test 
since we find a reduced $\chi^{2}$ of 1.0 for 25 dof.

\begin{figure}
\includegraphics[scale=0.5]{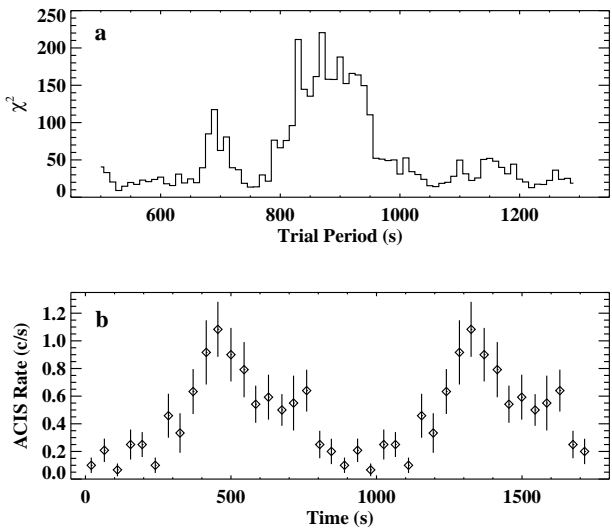}
\caption{Results of folding the {\em Chandra} light curve for CXOU~J180839.8--274131.  ({\em a}) The
resulting $\chi^{2}$ value when fitting the light curve with a constant after folding on trial periods
between 500\,s and 1300\,s.  ({\em b}) The folded light curve for a period of 870\,s, which is the
trial period that gave the maximum $\chi^{2}$.
\label{fig:lc_fold}}
\end{figure}

As CXOU~J140846.0--610754 and CXOU~J180839.8--274131 show evidence for variability, 
we performed light curve folding on many trial periods to determine if there is evidence
for periodic variability.  For each trial period, the $\chi^{2}$ for a fit with a constant
function is calculated, and a large value of $\chi^{2}$ can indicate periodic variability.
Although the former source did not show any candidate periods, CXOU~J180839.8--274131 
exhibits a broad and strong $\chi^{2}$ peak between 800\,s and 950\,s (Figure~\ref{fig:lc_fold}).  
Although we note that the entire duration of our observation only covers five periods of 
the candidate oscillation, the folded light curve (Figure~\ref{fig:lc_fold}) shows that 
if the signal is real, it has a very large amplitude (defined as maximum minus minimum divided 
by maximum plus minimum) of $\approx$80\%.

\subsection{IR Identifications}

\begin{figure*}
\begin{center}
\includegraphics[scale=0.9]{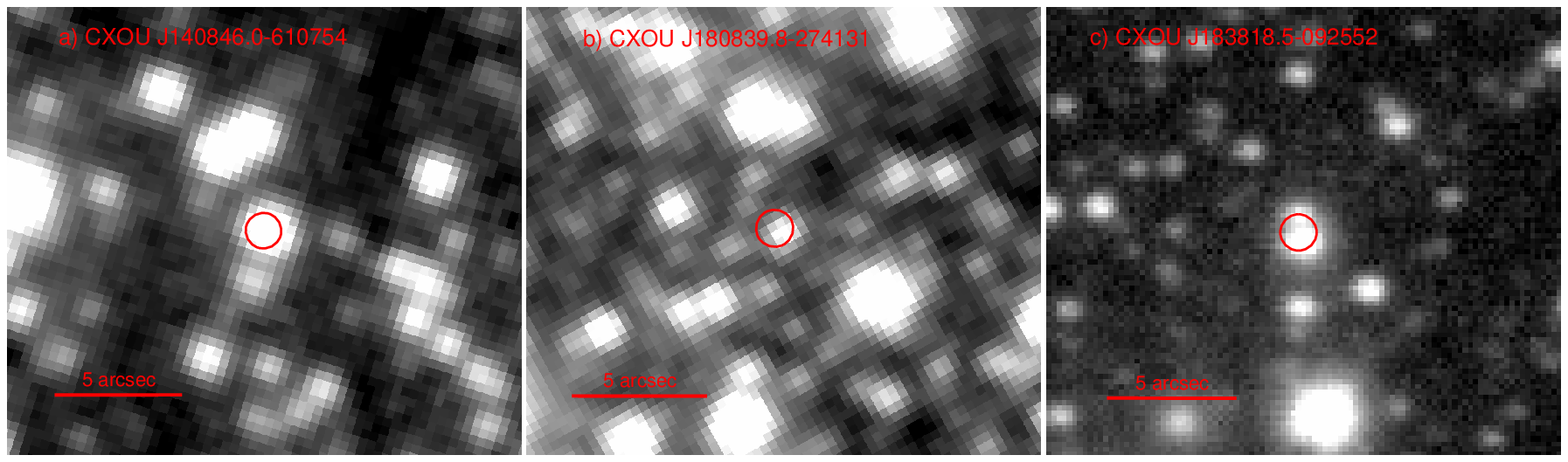}
\end{center}
\caption{({\em a}) $K_{\rm s}$-band image of CXOU~J140846.0--610754 from VISTA/VVV, ({\em b}) 
$K_{\rm s}$-band image of CXOU~J180839.8--274131 from VISTA/VVV, and ({\em c}) $K$-band image
of CXOU~J183818.5--092552 from UKIDSS.  The {\em Chandra} error circle is shown.
\label{fig:ir_images}}
\end{figure*}

The fields of CXOU~J140846.0--610754 and CXOU~J180839.8--274131 were covered by the VISTA 
Variables in the Via Lactea (VVV) survey \citep{minniti10}, and the field of CXOU~J183818.5--092552 
was covered by the UKIRT Infrared Deep Sky Survey (UKIDSS) survey \citep{lawrence07}.
We used the images from these surveys to identify possible near-IR counterparts.  
Figure~\ref{fig:ir_images}a shows the $K_{\rm s}$-band images with the {\em Chandra} 
positions.  For CXOU~J140846.0--610754, we confirm the association with VISTA source 
515845105705 (VVV~J140845.99--610754.1) reported in \cite{tomsick14_atel}, and we 
provide the VISTA $Z$-, $Y$-, $J$-, $H$-, and $K_{\rm s}$-magnitudes in Table~\ref{tab:mags}.  
As reported in \cite{tomsick14_atel}, this source is also listed in the {\em Spitzer}/GLIMPSE 
catalog as G312.1289+00.3516.

\begin{figure}
\includegraphics[scale=0.5]{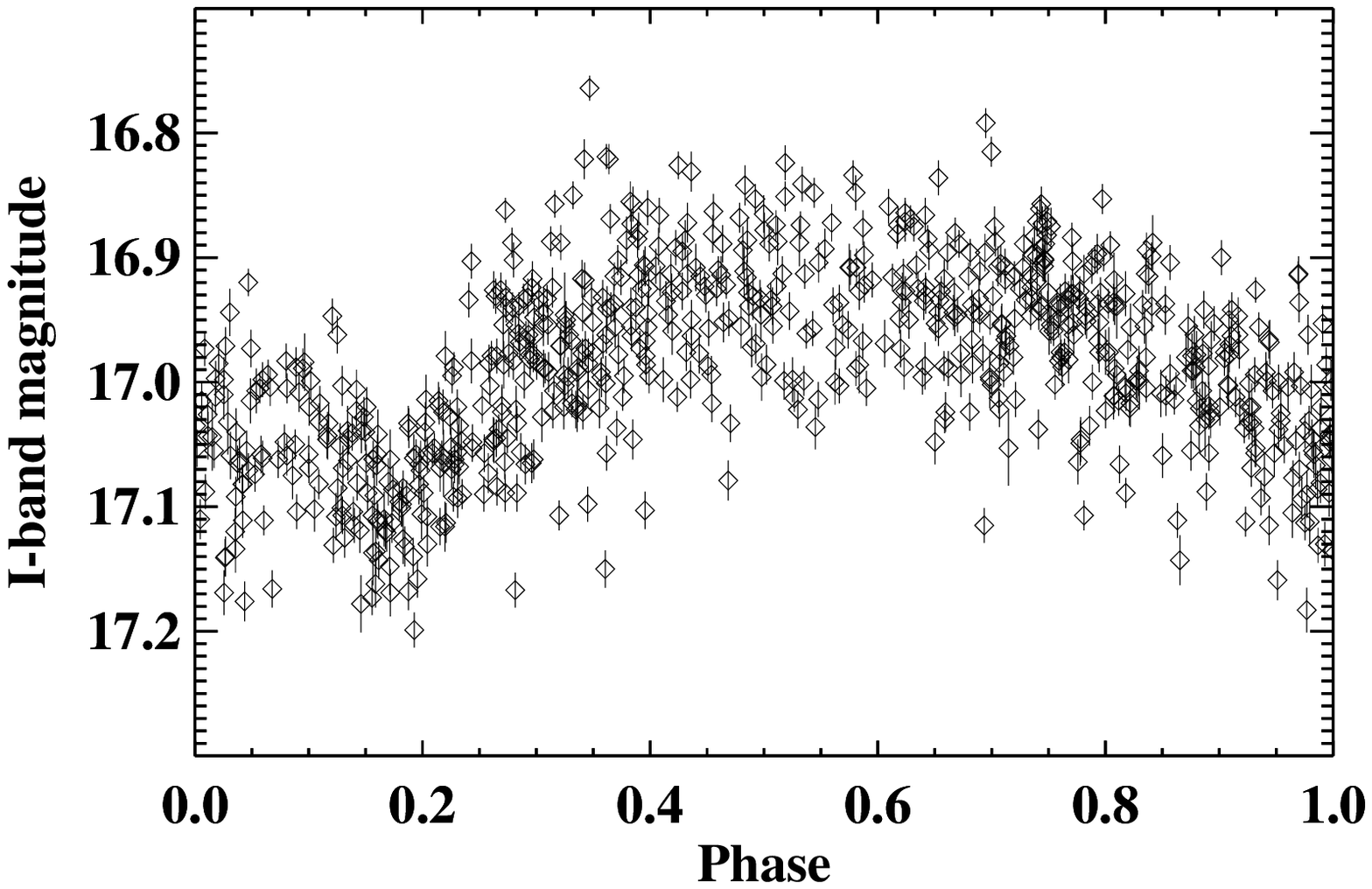}
\caption{OGLE $I$-band light curve for VVV~J180839.77--274131.7 after folding on the 0.28494563 day
period listed in the OGLE-III Catalog of Variable Stars \citep{soszynski11}.\label{fig:lc_ogle}}
\end{figure}

CXOU~J180839.8--274131 is within $0.\!^{\prime\prime}41$ of the VISTA source VVV~J180839.77--274131.7, 
and the $K_{\rm s}$-band image shown in Figure~\ref{fig:ir_images}b indicates that this is 
the only possible counterpart in the VVV images.  The VISTA magnitudes for this counterpart
are given in Table~\ref{tab:mags}.  We also used the VizieR website to search for possible
counterparts of this source in other catalogs, and we find a match with OGLE-BLG-RRLYR-14363, 
which is source number 14363 in the Optical Gravitational Lensing Experiment (OGLE-III) Catalog 
of Variable Stars \citep{soszynski11}.  The OGLE-III catalog indicates that the star varies 
periodically in the $I$-band with a period of $0.28494563\pm 0.00000062$ days.  We obtained the 
$I$-band measurements folded on this period, and the folded light curve is shown in 
Figure~\ref{fig:lc_ogle}.  While \cite{soszynski11} classify this source as an RR Lyrae star, 
a remark in the catalog indicates that the classification is uncertain, and we discuss this 
further in Section 4.

The position of CXOU~J183818.5--092552 is consistent with two near-IR sources from the
UKIDSS Galactic Plane Survey (UGPS):
UGPS~J183818.59--092551.8 at an angular distance of $0.\!^{\prime\prime}47$ and 
UGPS~J183818.58--092552.9 at an angular distance of $0.\!^{\prime\prime}73$.  Based on their
spatial profiles in the near-IR, both sources are classified as galaxies in the UGPS catalog.
The probability that the sources are stellar (i.e., not extended) is $<$1\% for 
UGPS~J183818.59--092551.8 and 5\% for UGPS~J183818.58--092552.9.  In the UGPS
catalog, the magnitudes listed for these sources do not appear to match the image shown in 
Figure~\ref{fig:ir_images}c.  Specifically, in the images, the source to the south 
(UGPS~J183818.58--092552.9) is clearly the brighter of the two, but the catalog lists it as 
$K = 13.040\pm 0.003$ (UKIDSS data release 10) compared to $K = 12.822\pm 0.002$ for the
northern source (UGPS~J183818.59--092551.8).  The UGPS survey has $K$-band images taken 
on 2005 September 28 and nearly three years later on 2008 May 31, and we examined both of 
the images to look for source variability; however, there are no discernible differences
between the two images.  The UKIDSS photometry is carried out using $2^{\prime\prime}$ 
radius extraction regions, and we suspect that this does not yield correct results for
these two sources, which have a separation of $\approx$$1^{\prime\prime}$.  We note that 
UGPS~J183818.59--092551.8 appears to be approximately the same brightness as the two sources 
to the south of the blended pair, and these are both listed as having $K = 14.2$ in the 
UKIDSS catalog.  Thus, we conclude that the true brightnesses of UGPS~J183818.59--092551.8 
and UGPS~J183818.58--092552.9 are likely $K\approx 14$ and $K\approx 13$, respectively 
(Table~\ref{tab:mags}).  The {\em Chandra} position is compatible with both sources, and both 
remain as candidate counterparts to CXOU~J183818.5--092552.

\subsection{Upper Limits for the Other Five IGR Sources}

For IGR~J15335--5420, IGR~J17164--3803, IGR~J17174--2436, IGR~J17306--2015, and IGR~J20107+4534, 
none of the {\em Chandra} sources listed in the table in the Appendix are clear counterparts, 
but these sources are possible counterparts.  We identified the highest count rate source in 
each field and took this as the upper limit on the {\em Chandra} count rate.  Then, extrapolating 
using a power-law model and the ACIS response matrix, we compared the {\em Chandra} count rate 
upper limit to the actual flux measurement for each of the five sources reported in \cite{krivonos12}.  
There are at least three reasons why the sources might be relatively faint in the {\em Chandra} 
energy band: 1. the source may be intrinsically hard; 2. the source may be highly absorbed; or 
3. the source may be variable (and any combination of these three is also possible).  
Table~\ref{tab:ul} lists the {\em Chandra} count rate upper limit and the 17--60\,keV flux 
measured by {\em INTEGRAL} \citep{krivonos12}.  Then, we performed calculations to consider the 
three effects mentioned above.  In the first calculation, we determined the values of $\Gamma$
required to produce the low {\em Chandra} count rate if the sources are unabsorbed.  Second, 
we assumed a value of $\Gamma = 2.1$, which is used in \cite{krivonos12} to determine
the {\em INTEGRAL} fluxes, and calculated the column densities that would lead to the 
{\em Chandra} count rate upper limits.  Finally, we assumed $\Gamma = 1$ and 
$N_{\rm H} = 5\times 10^{22}$\,cm$^{-2}$ and calculated the {\em Chandra} count rates that would 
be predicted for the {\em INTEGRAL} fluxes.  In Table~\ref{tab:ul}, we report the upper limits
on $\Gamma$ (first calculation), $N_{\rm H}$ (second calculation), and the variability factor, 
which is the ratio of the predicted count rate to the upper limit on the count rate.

\section{Discussion}

The result of the {\em Chandra} observations is that we have identified X-ray counterparts to
IGR~J14091--6108, IGR~J18088--2741, and IGR~J18381--0924, and henceforth, we will use the
IGR names to refer to these {\em Chandra}/{\em INTEGRAL} sources.  For IGR~J14091--6108 and
IGR~J18088--2741, the {\em Chandra} positions provide unique near-IR identifications, and for
IGR~J18381--0924, there are two near-IR sources compatible with the {\em Chandra} position.
Here, we discuss the nature of the three sources and then the implications of the 
parameter limits on the other five sources.

\subsection{IGR~J14091--6108}

There is clear evidence that IGR~J14091--6108 is in our Galaxy.  The iron line detected in
its spectrum is not redshifted, and the low measured column density indicates that it is 
relatively nearby: the Galactic column density along this line of sight ($l = 312.13^{\circ}$, 
$b$ = +$0.35^{\circ}$) is $5.4\times 10^{22}$\,cm$^{-2}$, but we measure an upper limit of 
$<$$1.7\times 10^{22}$\,cm$^{-2}$.  Although we do not know its intrinsic near-IR spectral shape, 
the fact that the VISTA magnitudes do not indicate strong reddening is also consistent with a 
relatively small distance, probably less than a few kpc.  

The hard X-ray spectrum ($\Gamma = 0.6\pm 0.4$) favors the possibilities that
either the compact object is a magnetic white dwarf in a CV \citep{revnivtsev08} or that 
it is an accreting neutron star with a high, $\sim$$10^{12}$\,G, magnetic field \citep{coburn02}.  
Nearly all of the accreting black holes that we know of have photon indices softer than 
$\Gamma\approx 1.2$--1.4, even in their hard state \citep{mr06}, and neutron stars with 
weaker magnetic fields have even softer spectra.  An isolated neutron star, such as a magnetar, 
could have a hard spectrum, but the brightness of IGR~J14091--6108 in the near-IR requires a 
stellar companion.  A CV with a spectrum as hard as we observe would almost certainly belong 
to the class of Intermediate Polars \citep[IPs,][]{scaringi10}, 
where the white dwarf magnetic field is strong enough to truncate the accretion disk and
funnel material onto the magnetic poles.  In the neutron star case, nearly all high magnetic 
field neutron stars in binary systems are found in HMXBs.  Thus, 
it is likely that IGR~J14091--6108 is either a CV/IP or a neutron star HMXB.

The source's X-ray luminosity could provide a way to distinguish between the different hypotheses 
because CV/IPs are not known to exceed values of $\sim$$10^{35}$\,erg\,s$^{-1}$.  With the 
0.3--10\,keV unabsorbed flux being $(2.2^{+0.4}_{-0.3})\times 10^{-12}$\,erg\,cm$^{-2}$\,s$^{-1}$, 
the luminosity is $3\times 10^{32}$ $d_{1}^{2}$\,erg\,s$^{-1}$, where $d_{1}$ is the distance 
in kpc, which leaves both possibilities open since neutron star HMXBs can have low 
luminosities.  The CV hypothesis is somewhat favored by the fact that neutron star HMXBs 
typically have iron line equivalent widths of 200\,eV or less \citep{gg15} while we measure
$EW = 1.2^{+0.9}_{-0.7}$\,keV.  Although there are some HMXBs with unusually high iron line strengths 
\citep{barragan09,gg15}, these only occur in cases where the continuum is absorbed by material 
intrinsic to the system (e.g., a stellar wind), but the low $N_{\rm H}$ for IGR~J14091--6108 makes 
this possibility unlikely.  While an $EW$ of 1.2\,keV is also somewhat high for a CV, values 
in excess of 400\,eV (combining all iron ionization states) are not unusual \citep{hm04}.
An even higher equivalent width for the iron lines was reported for a combined spectrum
consisting of {\em Chandra} sources in the Galactic Center region \citep{muno04}, and those
spectra also show evidence for emission lines from argon and calcium in the 3--4\,keV range.
As discussed in section 3.2, the IGR~J14091--6108 spectrum shows positive residuals at 
4\,keV, but they are not statistically significant.  \cite{muno04} interpret the Galactic 
Center point source spectrum as being due to a thermal plasma model, and they suggest that 
the emission is from a population of CV/IPs.

\subsection{IGR~J18088--2741}

A key piece of evidence that sheds light on the nature of IGR~J18088--2741 is its identification 
with OGLE-BLG-RRLYR-14363, which has a 6.84\,hr optical period.  Although it is contained in a 
catalog of RR~Lyrae stars \citep{soszynski11}, which produce periodic optical variations due to 
stellar pulsations, the fact that we detect this source with such a hard X-ray spectrum 
($\Gamma$ = --$0.7^{+0.4}_{-0.3}$) eliminates the possibility that the emission is coming from an 
isolated star.  The combination of the hard spectrum and the optical periodicity strongly suggests 
that we are seeing a binary system with a compact object, but the origin of the periodicity and 
the type of compact object are uncertain.  In many accreting systems, an optical modulation is 
seen at the orbital period of the system due to the heating of one side of the optical companion.  
However, such light curves are often approximately sinusoidal, and the fact that the IGR~J18088--2741 
optical light curve is more complex (see Figure~\ref{fig:lc_ogle}) may indicate that we are also 
seeing light from the accretion disk.  We also considered the possibility that we are seeing the
superhump period, which is related to tidal stresses in the accretion disks of CVs, but these 
are much more common for CVs with orbital periods less than a few hours \citep{patterson05}. 

Similarly to IGR~J14091--6108, the very hard X-ray spectrum of IGR~J18088--2741 favors the
possibilities that either the source is a CV/IP or that it is a highly magnetized neutron star 
in an X-ray binary.  Although the 800--950\,s X-ray period needs to be confirmed with a longer 
observation, by itself, such a period does not immediately distinguish between the CV/IP and X-ray 
binary possibilities.  However, the X-ray binary hypothesis is challenged by the fact that such 
slow rotators are only common in HMXBs, while, if the 6.84\,hr period is orbital, then this would 
be extremely short for an HMXB, where periods of days to weeks, or even months, are typical.

Given that its 0.3--10\,keV unabsorbed flux is $3\times 10^{-12}$\,erg\,cm$^{-2}$\,s$^{-1}$, this
source would need to lie at a distance in excess of 15\,kpc for its luminosity to challenge the CV/IP
limit of $\sim$$10^{35}$\,erg\,s$^{-1}$.  As the Galactic column density is $3\times 10^{21}$\,cm$^{-2}$ 
along this line of sight ($l = 3.65^{\circ}$, $b$ = --$3.84^{\circ}$), the X-ray measured $N_{\rm H}$ 
(see Table~\ref{tab:parameters}) does not provide any indication of the distance.  However, the fact 
that the optical extinction is low (see Table~\ref{tab:mags}) favors a much smaller distance than 15\,kpc, 
and with the luminosity being $4\times 10^{32}$ $d_{1}^{2}$\,erg\,s$^{-1}$, it is clear that this 
constraint does not rule out either possibility.

\subsection{IGR~J18381--0924}

The X-ray spectrum of IGR~J18381--0924 is significantly different from the other two sources.  The 
power-law is consistent with being softer, $\Gamma = 1.5^{+0.5}_{-0.4}$ (for the {\em Chandra}-only 
fit), and there is no evidence for a cutoff up to 86\,keV.  Also, the spectrum is absorbed, and the 
measured $N_{\rm H} = (4\pm 1)\times 10^{22}$\,cm$^{-2}$ is marginally higher than the Galactic column 
density along this line of sight ($l = 23.05^{\circ}$, $b$ = --$1.36^{\circ}$) of $2.1\times 10^{22}$\,cm$^{-2}$.  
Based on just the X-ray spectrum, the source could be Galactic or extragalactic:  if it is Galactic, 
then an accreting black hole in the hard state is a possibility; and, if it is extragalactic, then its 
properties are consistent with those of AGN.  Both types of sources might have radio counterparts, 
but there is no radio counterpart listed in the NRAO VLA Sky Survey \citep[NVSS,][]{condon98} even 
though this area of the sky was covered.  The interpretation of this source as being Galactic starts 
to become problematic when one considers the lack of X-ray variability on long and short time scales 
since most Galactic black holes are X-ray transients.  Even stronger evidence in favor of an extragalactic 
nature comes from the fact that the two possible near-IR counterparts are identified as galaxies.  
Thus, we conclude that IGR~J18381--0924 is probably an AGN.

\subsection{Implications of the Upper Limits for the Other Five Fields}

For the sources without clear {\em Chandra} counterparts (IGR~J15335--5420, IGR~J17164--3803, 
IGR~J17174--2436, IGR~J17306--2015, and IGR~J20107+4534), Table~\ref{tab:ul} shows that the 
sources need to be extremely hard or highly variable in order to explain the {\em Chandra}
upper limits.  Values of $\Gamma$ = --1.2 to --1.5 may be possible for some types of 
Galactic sources (e.g., IGR~J18088--2741), but $\Gamma$ = --2.9 is too hard to be realistic.
Explaining the low {\em Chandra} count rates by high column densities requires value of
$N_{\rm H}$ of $\sim$$2\times 10^{24}$\,cm$^{-2}$.  Such values occur very rarely in Galactic
sources but have been seen for the most extreme obscured HMXB \citep{mg03,barragan09}.  
Such column densities occur relatively often for Seyfert 2 AGN \citep[][and references therein]{brightman15}, 
and we consider this a possibility for the nature of some of these sources.  On the other
hand, the low {\em Chandra} count rates can be explained by source variability with factors
of $>$11--60.  Such variability is possible for many types of Galactic sources.  The observations 
are consistent with either long-term (years to decades) or short-term (hours) variability.

\section{Conclusions and Future Work}

We conclude that IGR~J14091--6108 and IGR~J18088--2741 may be either CVs or HMXBs but that a
CV is favored for the former source by the strong iron line, and a CV is strongly favored in
the latter source if the 6.84\,hr optical periodicity is the orbital period of the system.  
If the CV nature is confirmed by optical or near-IR spectroscopy, then these sources would
join a group of 26 IGR CVs.  An AGN nature is most likely for IGR~J18381--0924.  Follow-up 
optical or IR spectroscopy of the two candidate near-IR counterparts, which are both identified 
as galaxies in the UGPS catalog, is required to confirm that one or both of the counterparts
are AGN.  For the other five IGR sources, the fact that we were not able to definitively 
determine a unique counterpart using the {\em Chandra} observations may indicate that these
IGR sources are transient, highly variable, or that their spectra are extremely hard
(either due to absorption or intrinsic spectral shape).  Further observations of the sources
listed in the Appendix may still provide a determination of the correct counterpart and
the nature of these sources.

\acknowledgments

JAT and QW acknowledge partial support from the National Aeronautics and Space Administration
(NASA) through {\em Chandra} Award Number GO4-15044X issued by the {\em Chandra} X-ray Observatory 
Center, which is operated by the Smithsonian Astrophysical Observatory under NASA contract NAS8-03060.  
RK acknowledges support from Russian Science Foundation (grant 14-22-00271).  This work included 
observations with {\em INTEGRAL}, an ESA project with instruments and the science data center 
funded by ESA member states (especially the PI countries: Denmark, France, Germany, Italy, 
Switzerland, Spain), and Poland, and with the participation of Russia and the USA.  This research 
has made use of the IGR Sources page maintained by J.~Rodriguez and A.~Bodaghee 
(http://irfu.cea.fr/Sap/IGR-Sources), the VizieR catalog access tool and the SIMBAD database, which 
are both operated at CDS, Strasbourg, France.  This work is based in part on data obtained as part 
of the UKIRT Infrared Deep Sky Survey and also on data products from the VVV Survey observations 
made with the VISTA telescope at the ESO Paranal Observatory under program ID 179.B-2002.  This 
publication makes use of data products from the Two Micron All Sky Survey, which is a joint 
project of the University of Massachusetts and the Infrared Processing and Analysis Center/California 
Institute of Technology, funded by NASA and the National Science Foundation.  We acknowledge 
suggestions from an anonymous referee, which led to improvements in the manuscript.

\appendix

Section 3.1 describes how we used {\ttfamily wavdetect} to find 115 {\em Chandra}
sources in the eight ACIS-I fields.  A full source lists is provided in 
Table~\ref{tab:sourcelist}, and here we provide a more thorough explanation of how 
the values in those tables were derived.  The sources are listed in order of how far 
they lie from the centers of each of the {\em INTEGRAL} error circles, and this angle 
($\theta$), in arcminutes, is given in the tables.  

To determine the source positions, we started with the values provided by 
{\ttfamily wavdetect}.  In order to reduce the systematic uncertainties due to the 
absolute {\em Chandra} pointing, we looked for matches to the {\em Chandra} sources in 
near-IR catalogs.  We used the VISTA Variables in the Via Lactea (VVV) near-IR catalog 
\citep{minniti10} for sources in the IGR~J14091--6108, IGR~J15335--5420, IGR~J17164--3803, 
and IGR~J18088--2741 fields.  Due to the incomplete sky coverage of the VVV catalog, 
counterparts for other fields were not found in VVV.  Thus, for IGR~J17174--2436 and 
IGR~J18381--0924, we searched the UKIRT Infrared Deep Sky Survey (UKIDSS) catalog 
\citep{lawrence07}, and for IGR~J17306--2015 and IGR~J20107+4534, we searched the 
2 Micron All Sky Survey (2MASS) catalog \citep{cutri03}.  For each field, we 
compiled a list of potential near-IR counterparts, and ran {\ttfamily wcs\_match}
to compare the {\em Chandra} and near-IR lists, determine likely matches, and 
calculate the astrometric correction.  When we ran {\ttfamily wcs\_match}, we used 
the ``residlim'' parameter to only include sources for which the {\em Chandra} and 
near-IR positions agree to better than $1^{\prime\prime}$ in order to discard unlikely
matches as well as many of the {\em Chandra} sources with larger statistical 
position errors.  We only found one match in the IGR~J17174--2436, two matches in the
IGR~J17306--2015 field, and three matches in the IGR~J18381--0924 field, which are not 
enough to calculate a reliable position shift.  For the other five fields, we found 
between 5 and 15 matches, and the number of matches for each field is given in 
Table~\ref{tab:matches_and_shifts}.  Although {\ttfamily wcs\_match} can consider both 
linear translations and coordinate rotations to optimize source matches, we restricted 
the shifts to linear translations since there are three fields where we only have five 
matches.  The shifts determined by {\ttfamily wcs\_match} are given in 
Table~\ref{tab:matches_and_shifts}.  We used {\ttfamily wcs\_update} to 
apply these shifts to the positions reported in Table~\ref{tab:sourcelist}.

The position uncertainties have statistical and systematic contributions.  Since the
{\em Chandra} PSF becomes significantly larger for off-axis angles, the statistical
uncertainty depends on the number of counts detected for each source as well as its
off-axis angle.  We used Equation 5 from \cite{hong05} to calculate the statistical
uncertainty.  Although this equation gives a 95\% confidence error circle radius
(2-$\sigma$), we reduced the radius to 90\% confidence (1.7-$\sigma$) by multiplying
by a factor of 1.7/2.0.  For the systematic contribution, we assumed a 90\% confidence
error of $0.\!^{\prime\prime}64$ \citep{weisskopf05}.  While this is an overestimate 
for the fields where we shifted coordinates, we could not determine how much these
shifts decrease the systematic position uncertainty due to the relatively small number 
of X-ray/near-IR matches.  The total position uncertainty quoted in the tables is the
value obtained by adding the statistical and systematic values in quadrature.



\clearpage

\begin{table}
\caption{{\em Chandra} Observations\label{tab:obs}}
\begin{minipage}{\linewidth}
\begin{center}
\footnotesize
\begin{tabular}{cccclcc} \hline \hline
IGR Name & ObsID & $l$\footnote{Galactic longitude measured by {\em INTEGRAL} (degrees)} & $b$\footnote{Galactic latitude measured by {\em INTEGRAL} (degrees)} & Start Time & Exposure Time (s) & $N_{\rm sources}$\footnote{Number of {\em Chandra} sources detected in each field.}\\ \hline \hline
J14091--6108 & 15789 & 312.13 &  +0.35 & 2013 Dec  7, 23.7 h UT & 4912 & 18\\ 
J15335--5420 & 15790 & 325.18 &  +1.36 & 2014 Jun  4,  8.9 h UT & 4912 & 8\\
J17164--3803 & 15788 & 349.07 &  +0.07 & 2014 Apr 15, 18.6 h UT & 4912 & 12\\
J17174--2436 & 15797 &   0.19 &  +7.65 & 2014 Feb 27,  5.3 h UT & 4902 & 14\\
J17306--2015 & 15796 &   5.52 &  +7.53 & 2015 Feb  8, 11.2 h UT & 4906 & 7\\
J18088--2741 & 15794 &   3.65 & --3.84 & 2014 Feb 16, 16.7 h UT & 4915 & 29\\
J18381--0924 & 15791 &  23.05 & --1.36 & 2014 Feb 15, 18.6 h UT & 4912 & 14\\
J20107+4534  & 15795 &  81.39 &  +6.59 & 2013 Dec 21,  5.4 h UT & 4912 & 13\\ \hline
\end{tabular}
\end{center}
\end{minipage}
\end{table}

\begin{table}
\caption{Spectral Parameters\label{tab:parameters}}
\begin{minipage}{\linewidth}
\begin{center}
\footnotesize
\begin{tabular}{ccccc} \hline \hline
Parameter\footnote{The errors on the parameters are 90\% confidence.} & Units & {\em Chandra}-only & \multicolumn{2}{c}{{\em Chandra}+{\em INTEGRAL}}\\ \hline\hline
\multicolumn{5}{c}{CXOU~J140846.0--610754}\\ \hline
$N_{\rm H}$\footnote{The column density is calculated assuming \cite{wam00} abundances and \cite{vern96} cross sections.  The Galactic column density (atomic and molecular Hydrogen) at the locations of CXOU~J140846.0--610754, CXOU~J180839.8--274131, and CXOU~J183818.5--092552 are $5.4\times 10^{22}$\,cm$^{-2}$, $3.0\times 10^{21}$\,cm$^{-2}$, and $2.1\times 10^{22}$\,cm$^{-2}$, respectively \citep{kalberla05,dht01}.} 
           & $10^{22}$\,cm$^{-2}$ & $1.0^{+0.7}_{-0.6}$     & $0.8^{+0.7}_{-0.5}$     & $0.6^{+0.6}_{-0.5}$\\
$\Gamma$   &    --              & $0.6\pm 0.4$          & $0.3\pm 0.5$          & $0.1^{+0.4}_{-0.5}$\\
Unabsorbed Flux (0.3--10\,keV) & $10^{-12}$\,erg\,cm$^{-2}$\,s$^{-1}$ 
                                & $2.2^{+0.4}_{-0.3}$     & $2.3\pm 0.3$          & $2.4\pm 0.3$\\
$E_{\rm line}$  &   keV           & $6.6\pm 0.2$          & $6.6^{+0.4}_{-0.2}$     & $6.6^{+0.3}_{-0.2}$\\
$\sigma_{\rm line}$ & keV         & $0.2^{+0.4}_{-0.2}$     & $<$0.6                & $<$0.6\\
$N_{\rm line}$ & $10^{-5}$\,ph\,cm$^{-2}$\,s$^{-1}$
                                & $3.0^{+2.3}_{-1.8}$     & $3.4^{+3.6}_{-2.0}$     & $3.0^{+1.9}_{-1.8}$\\
$E_{\rm fold}$ &  keV             & --                    & $13^{+10}_{-5}$        & $14^{+11}_{-5}$\\
$N_{INTEGRAL}/N_{Chandra}$\footnote{The {\em INTEGRAL} normalization relative to {\em Chandra}.} &   --   
                                &   --                  & $1.8^{+2.8}_{-1.1}$     & 1.0\footnote{Fixed.}\\
$\chi^{2}$/dof   &   --          & 18.9/19               & 18.5/21              & 19.7/22\\ \hline
\multicolumn{5}{c}{CXOU~J180839.8--274131}\\ \hline
$N_{\rm H}$$^{b}$ & $10^{22}$\,cm$^{-2}$ &    $0.2^{+1.0}_{-0.2}$   & $<$0.7             & $<$0.7\\
$\Gamma$   &    --                    & --$0.7^{+0.4}_{-0.3}$   & --$1.5\pm 0.4$     & --$1.5^{+0.4}_{-0.2}$\\
Unabsorbed Flux (0.3--10\,keV) & $10^{-12}$\,erg\,cm$^{-2}$\,s$^{-1}$   
                                &   $3.5^{+0.7}_{-0.6}$         & $2.8\pm 0.5$       & $2.8\pm 0.4$\\
$E_{\rm fold}$ &  keV             & --                         & $5.1^{+1.6}_{-1.3}$   & $4.8^{+1.1}_{-0.8}$\\
$N_{INTEGRAL}/N_{Chandra}$$^{c}$     &   --   
                                &   --                       & $0.8^{+0.6}_{-0.4}$    & 1.0$^{d}$\\
$\chi^{2}$/dof   &   --          &  13.6/15                   & 12.0/17             & 12.3/18\\ \hline
\multicolumn{5}{c}{CXOU~J183818.5--092552}\\ \hline
$N_{\rm H}$$^{b}$ & $10^{22}$\,cm$^{-2}$  & $4.0^{+1.3}_{-1.2}$      & $4.3^{+1.4}_{-1.1}$ & $3.8^{+0.9}_{-0.7}$\\
$\Gamma$   &    --              & $1.5^{+0.5}_{-0.4}$      & $1.6\pm 0.4$      & $1.4\pm 0.1$\\
Unabsorbed Flux (0.3--10\,keV)  & $10^{-12}$\,erg\,cm$^{-2}$\,s$^{-1}$   
                                & $3.4^{+1.4}_{-0.6}$      & $3.5^{+1.4}_{-0.7}$  & $3.2\pm 0.5$\\
$N_{INTEGRAL}/N_{Chandra}$$^{c}$     &   --   
                                &   --                  &  $1.6^{+2.1}_{-0.9}$ & 1.0$^{d}$\\
$\chi^{2}$/dof   &   --          &  14.4/22              &  20.4/25          & 21.2/26\\ \hline
\end{tabular}
\end{center}
\end{minipage}
\end{table}

\begin{table}
\caption{IR Identifications\label{tab:mags}}
\begin{minipage}{\linewidth}
\begin{center}
\footnotesize
\begin{tabular}{ccc} \hline \hline
IR Source & Separation\footnote{The angular separation between the {\em Chandra} position and the catalog position.} & Magnitudes\\ \hline \hline
\multicolumn{3}{c}{CXOU J140846.0--610754}\\ \hline
VVV J140845.99--610754.1 & $0.\!^{\prime\prime}16$ & $Z = 16.165\pm 0.005$\\
          ''             &          ''          & $Y = 15.806\pm 0.005$\\
          ''             &          ''          & $J = 15.220\pm 0.005$\\
          ''             &          ''          & $H = 14.833\pm 0.007$\\
          ''             &          ''          & $K_{\rm s} = 14.394\pm 0.010$\\ \hline
\multicolumn{3}{c}{CXOU J180839.8--274131}\\ \hline
VVV J180839.77--274131.7 & $0.\!^{\prime\prime}41$ & $Z = 16.77\pm 0.06$\\
          ''             &          ''          & $Y = 16.65\pm 0.07$\\
          ''             &          ''          & $J = 16.09\pm 0.06$\\
          ''             &          ''          & $H = 15.82\pm 0.08$\\
          ''             &          ''          & $K_{\rm s} = 15.67\pm 0.09$\\ \hline
\multicolumn{3}{c}{CXOU J183818.5--092552}\\ \hline
UGPS J183818.59--092551.8 & $0.\!^{\prime\prime}47$ & $K\approx 14$\footnote{As described in the text, the magnitudes listed in the UKIDSS catalog are not consistent with the images.  Here, we provide estimates for the $K$-band magnitudes based on visual comparisons to other nearby sources.}\\
UGPS J183818.58--092552.9 & $0.\!^{\prime\prime}73$ & $K\approx 13$\\ \hline
\end{tabular}
\end{center}
\end{minipage}
\end{table}

\begin{table}
\caption{Limits on Parameters for the IGR Sources without Clear Counterparts\label{tab:ul}}
\begin{minipage}{\linewidth}
\begin{center}
\footnotesize
\begin{tabular}{cccccc} \hline \hline
IGR Name & Limit on ACIS & {\em INTEGRAL}  & Limit on & Limit on & Limit on\\
         & Rate (c/s)    & Flux\footnote{The 17--60\,keV flux in units of erg\,cm$^{-2}$\,s$^{-1}$ from \cite{krivonos12}.} & $\Gamma$ & $N_{\rm H}$ (cm$^{-2}$) & Variability\footnote{The ratio of the {\em Chandra} count rate assuming the {\em INTEGRAL} flux, $\Gamma = 1$, and $N_{\rm H} = 5\times 10^{22}$\,cm$^{-2}$ to the measured upper limit on the count rate.}\\ \hline \hline
J15335--5420 & $<$0.0034 & $5.5\times 10^{-12}$ & $<$--1.5 & $>$$1.80\times 10^{24}$ & $>$16\\
J17164--3803 & $<$0.0057 & $8.7\times 10^{-12}$ & $<$--1.5 & $>$$1.75\times 10^{24}$ & $>$15\\
J17174--2436 & $<$0.0063 & $6.4\times 10^{-12}$ & $<$--1.2 & $>$$1.50\times 10^{24}$ & $>$10\\
J17306--2015 & $<$0.0025 & $1.6\times 10^{-11}$ & $<$--2.9 & $>$$2.85\times 10^{24}$ & $>$60\\
J20107+4534  & $<$0.0062 & $6.8\times 10^{-12}$ & $<$--1.2 & $>$$1.55\times 10^{24}$ & $>$11\\ \hline
\end{tabular}
\end{center}
\end{minipage}
\end{table}

{\footnotesize
\begin{center}
\begin{longtable}{ccccccc}
\caption{{\em Chandra} Sources in IGR Source Fields\label{tab:sourcelist}} \\
\hline 
\multicolumn{1}{c}{Source} & 
\multicolumn{1}{c}{$\theta$$^{a}$} & 
\multicolumn{1}{c}{{\em Chandra} R.A.} & 
\multicolumn{1}{c}{{\em Chandra} Decl.} & 
\multicolumn{1}{c}{Position}  & 
\multicolumn{1}{c}{ACIS} &  
\multicolumn{1}{c}{} \\
\multicolumn{1}{c}{Number} & 
\multicolumn{1}{c}{(arcminutes)} & 
\multicolumn{1}{c}{(J2000)}  & 
\multicolumn{1}{c}{(J2000)} & 
\multicolumn{1}{c}{Uncertainty$^{b}$} &
\multicolumn{1}{c}{Counts$^{c}$} & 
\multicolumn{1}{c}{Hardness$^{d}$}\\ \hline
\endfirsthead
\caption{Continued}\\
\hline
\multicolumn{1}{c}{Source} & 
\multicolumn{1}{c}{$\theta$$^{a}$} & 
\multicolumn{1}{c}{{\em Chandra} R.A.} & 
\multicolumn{1}{c}{{\em Chandra} Decl.} & 
\multicolumn{1}{c}{Position}  & 
\multicolumn{1}{c}{ACIS} &  
\multicolumn{1}{c}{} \\
\multicolumn{1}{c}{Number} & 
\multicolumn{1}{c}{(arcminutes)} & 
\multicolumn{1}{c}{(J2000)}  & 
\multicolumn{1}{c}{(J2000)} & 
\multicolumn{1}{c}{Uncertainty$^{b}$} &
\multicolumn{1}{c}{Counts$^{c}$} & 
\multicolumn{1}{c}{Hardness$^{d}$}\\ \hline
\endhead
\hline
\endfoot
\hline
\endlastfoot
\multicolumn{7}{c}{IGR J14091--6108}\\ \hline
1 & 0.45 & $14^{\rm h}08^{\rm m}46^{\rm s}.02$ & $-61^{\circ}07^{\prime}54.\!^{\prime\prime}2$ & $0.\!^{\prime\prime}69$ & $405^{+21}_{-20}$ & $0.56\pm0.06$ \\
2 & 1.13 & $14^{\rm h}08^{\rm m}36^{\rm s}.70$ & $-61^{\circ}06^{\prime}48.\!^{\prime\prime}1$ & $0.\!^{\prime\prime}86$ & $3.8^{+3.2}_{-1.9}$ & $<0.53$ \\
3 & 1.18 & $14^{\rm h}08^{\rm m}42^{\rm s}.96$ & $-61^{\circ}08^{\prime}47.\!^{\prime\prime}7$ & $0.\!^{\prime\prime}80$ & $5.8^{+3.6}_{-2.4}$ & $<-0.09$ \\
4 & 1.58 & $14^{\rm h}08^{\rm m}31^{\rm s}.37$ & $-61^{\circ}06^{\prime}56.\!^{\prime\prime}1$ & $0.\!^{\prime\prime}72$ & $43^{+8}_{-7}$ & $>0.70$ \\
5 & 1.59 & $14^{\rm h}08^{\rm m}30^{\rm s}.19$ & $-61^{\circ}07^{\prime}52.\!^{\prime\prime}2$ & $0.\!^{\prime\prime}90$ & $3.8^{+3.2}_{-1.9}$ & $>-0.27$ \\
6 & 2.08 & $14^{\rm h}08^{\rm m}27^{\rm s}.89$ & $-61^{\circ}08^{\prime}33.\!^{\prime\prime}5$ & $0.\!^{\prime\prime}91$ & $4.8^{+3.4}_{-2.2}$ & $<0.04$ \\
7 & 2.50 & $14^{\rm h}08^{\rm m}49^{\rm s}.80$ & $-61^{\circ}05^{\prime}14.\!^{\prime\prime}2$ & $0.\!^{\prime\prime}76$ & $23^{+6}_{-5}$ & $-0.05\pm0.28$ \\
8 & 2.66 & $14^{\rm h}08^{\rm m}34^{\rm s}.99$ & $-61^{\circ}05^{\prime}08.\!^{\prime\prime}6$ & $0.\!^{\prime\prime}78$ & $18^{+5}_{-4}$ & $<-0.40$ \\
9 & 3.87 & $14^{\rm h}08^{\rm m}52^{\rm s}.79$ & $-61^{\circ}11^{\prime}18.\!^{\prime\prime}4$ & $1.\!^{\prime\prime}41$ & $4.5^{+3.4}_{-2.2}$ & $<0.06$ \\
10 & 4.35 & $14^{\rm h}08^{\rm m}40^{\rm s}.33$ & $-61^{\circ}03^{\prime}16.\!^{\prime\prime}8$ & $1.\!^{\prime\prime}40$ & $6.0^{+3.8}_{-2.6}$ & $0.07\pm0.72$ \\
11 & 5.10 & $14^{\rm h}08^{\rm m}32^{\rm s}.36$ & $-61^{\circ}12^{\prime}32.\!^{\prime\prime}2$ & $1.\!^{\prime\prime}65$ & $6.8^{+4.0}_{-2.8}$ & $<-0.11$ \\
12 & 7.28 & $14^{\rm h}08^{\rm m}16^{\rm s}.64$ & $-61^{\circ}01^{\prime}04.\!^{\prime\prime}6$ & $3.\!^{\prime\prime}04$ & $8.0^{+4.6}_{-3.4}$ & $<-0.46$ \\
13 & 7.64 & $14^{\rm h}09^{\rm m}42^{\rm s}.67$ & $-61^{\circ}10^{\prime}15.\!^{\prime\prime}2$ & $2.\!^{\prime\prime}60$ & $11^{+5}_{-4}$ & $<-0.16$ \\
14 & 7.83 & $14^{\rm h}09^{\rm m}44^{\rm s}.69$ & $-61^{\circ}05^{\prime}08.\!^{\prime\prime}7$ & $2.\!^{\prime\prime}08$ & $17^{+6}_{-5}$ & $>0.22$ \\
15 & 8.32 & $14^{\rm h}08^{\rm m}13^{\rm s}.99$ & $-61^{\circ}15^{\prime}09.\!^{\prime\prime}3$ & $7.\!^{\prime\prime}18$ & $4.6^{+4.6}_{-3.5}$ & --- \\
16 & 8.66 & $14^{\rm h}09^{\rm m}04^{\rm s}.29$ & $-61^{\circ}15^{\prime}53.\!^{\prime\prime}8$ & $1.\!^{\prime\prime}23$ & $75^{+10}_{-9}$ & $>0.77$ \\
17 & 8.80 & $14^{\rm h}08^{\rm m}18^{\rm s}.16$ & $-61^{\circ}15^{\prime}52.\!^{\prime\prime}5$ & $2.\!^{\prime\prime}20$ & $23^{+7}_{-6}$ & $>0.56$ \\
18 & 9.28 & $14^{\rm h}09^{\rm m}59^{\rm s}.91$ & $-61^{\circ}08^{\prime}13.\!^{\prime\prime}7$ & $1.\!^{\prime\prime}94$ & $34^{+8}_{-7}$ & $0.13\pm0.24$ \\ \hline
\multicolumn{7}{c}{IGR J15335--5420}\\ \hline
1 & 0.65 & $15^{\rm h}33^{\rm m}34^{\rm s}.98$ & $-54^{\circ}21^{\prime}03.\!^{\prime\prime}2$ & $0.\!^{\prime\prime}80$ & $4.9^{+3.4}_{-2.2}$ & $<0.03$ \\
2 & 1.48 & $15^{\rm h}33^{\rm m}34^{\rm s}.87$ & $-54^{\circ}23^{\prime}03.\!^{\prime\prime}9$ & $0.\!^{\prime\prime}85$ & $4.8^{+3.4}_{-2.2}$ & $>-0.05$ \\
3 & 3.22 & $15^{\rm h}33^{\rm m}30^{\rm s}.98$ & $-54^{\circ}18^{\prime}24.\!^{\prime\prime}5$ & $1.\!^{\prime\prime}15$ & $4.7^{+3.4}_{-2.2}$ & $<0.30$ \\
4 & 4.31 & $15^{\rm h}33^{\rm m}39^{\rm s}.34$ & $-54^{\circ}17^{\prime}24.\!^{\prime\prime}8$ & $1.\!^{\prime\prime}35$ & $6.2^{+3.8}_{-2.6}$ & $>0.13$ \\
5 & 5.12 & $15^{\rm h}32^{\rm m}57^{\rm s}.89$ & $-54^{\circ}22^{\prime}14.\!^{\prime\prime}6$ & $1.\!^{\prime\prime}29$ & $11^{+5}_{-3}$ & $>0.14$ \\
6 & 5.25 & $15^{\rm h}33^{\rm m}04^{\rm s}.40$ & $-54^{\circ}18^{\prime}23.\!^{\prime\prime}1$ & $2.\!^{\prime\prime}22$ & $4.7^{+3.6}_{-2.4}$ & $<0.42$ \\
7 & 8.19 & $15^{\rm h}32^{\rm m}47^{\rm s}.81$ & $-54^{\circ}26^{\prime}32.\!^{\prime\prime}7$ & $2.\!^{\prime\prime}31$ & $17^{+6}_{-5}$ & $0.09\pm0.38$ \\
8 & 8.38 & $15^{\rm h}33^{\rm m}21^{\rm s}.52$ & $-54^{\circ}29^{\prime}50.\!^{\prime\prime}1$ & $4.\!^{\prime\prime}42$ & $7.9^{+4.9}_{-3.7}$ & $<0.17$ \\ \hline
\multicolumn{7}{c}{IGR J17164--3803}\\ \hline
1 & 0.17 & $17^{\rm h}16^{\rm m}29^{\rm s}.26$ & $-38^{\circ}02^{\prime}04.\!^{\prime\prime}9$ & $0.\!^{\prime\prime}75$ & $7.9^{+4.0}_{-2.8}$ & $<0.11$ \\
2 & 1.87 & $17^{\rm h}16^{\rm m}26^{\rm s}.97$ & $-38^{\circ}00^{\prime}07.\!^{\prime\prime}8$ & $0.\!^{\prime\prime}73$ & $28^{+6}_{-5}$ & $0.36\pm0.26$ \\
3 & 2.26 & $17^{\rm h}16^{\rm m}33^{\rm s}.23$ & $-37^{\circ}59^{\prime}46.\!^{\prime\prime}9$ & $0.\!^{\prime\prime}93$ & $4.8^{+3.4}_{-2.2}$ & $<0.30$ \\
4 & 2.89 & $17^{\rm h}16^{\rm m}19^{\rm s}.94$ & $-38^{\circ}04^{\prime}06.\!^{\prime\prime}1$ & $1.\!^{\prime\prime}35$ & $2.8^{+2.9}_{-1.7}$ & $<0.61$ \\
5 & 3.82 & $17^{\rm h}16^{\rm m}17^{\rm s}.33$ & $-37^{\circ}58^{\prime}58.\!^{\prime\prime}9$ & $0.\!^{\prime\prime}94$ & $12^{+5}_{-4}$ & $>0.22$ \\
6 & 4.01 & $17^{\rm h}16^{\rm m}09^{\rm s}.39$ & $-38^{\circ}02^{\prime}22.\!^{\prime\prime}6$ & $1.\!^{\prime\prime}23$ & $6.4^{+3.8}_{-2.6}$ & $>-0.28$ \\
7 & 4.34 & $17^{\rm h}16^{\rm m}49^{\rm s}.50$ & $-38^{\circ}03^{\prime}48.\!^{\prime\prime}3$ & $1.\!^{\prime\prime}67$ & $4.4^{+3.4}_{-2.2}$ & $<0.09$ \\
8 & 5.83 & $17^{\rm h}16^{\rm m}02^{\rm s}.06$ & $-38^{\circ}04^{\prime}02.\!^{\prime\prime}9$ & $1.\!^{\prime\prime}37$ & $14^{+5}_{-4}$ & $0.05\pm0.40$ \\
9 & 6.57 & $17^{\rm h}17^{\rm m}00^{\rm s}.25$ & $-37^{\circ}59^{\prime}19.\!^{\prime\prime}0$ & $2.\!^{\prime\prime}96$ & $6.1^{+4.1}_{-3.0}$ & $<0.04$ \\
10 & 6.87 & $17^{\rm h}16^{\rm m}02^{\rm s}.99$ & $-38^{\circ}06^{\prime}22.\!^{\prime\prime}2$ & $1.\!^{\prime\prime}97$ & $12^{+5}_{-4}$ & $0.17\pm0.46$ \\
11 & 7.17 & $17^{\rm h}16^{\rm m}58^{\rm s}.28$ & $-37^{\circ}57^{\prime}30.\!^{\prime\prime}7$ & $1.\!^{\prime\prime}54$ & $21^{+6}_{-5}$ & $>0.35$ \\
12 & 7.29 & $17^{\rm h}16^{\rm m}56^{\rm s}.86$ & $-38^{\circ}06^{\prime}52.\!^{\prime\prime}0$ & $1.\!^{\prime\prime}52$ & $23^{+6}_{-5}$ & $0.13\pm0.29$ \\ \hline
\multicolumn{7}{c}{IGR J17174--2436}\\ \hline
1 & 1.92 & $17^{\rm h}17^{\rm m}20^{\rm s}.44$ & $-24^{\circ}33^{\prime}58.\!^{\prime\prime}4$ & $0.\!^{\prime\prime}83$ & $6.8^{+3.8}_{-2.6}$ & $<0.22$ \\
2 & 2.75 & $17^{\rm h}17^{\rm m}10^{\rm s}.65$ & $-24^{\circ}34^{\prime}06.\!^{\prime\prime}6$ & $0.\!^{\prime\prime}89$ & $7.7^{+4.0}_{-2.8}$ & $-0.02\pm0.58$ \\
3 & 2.92 & $17^{\rm h}17^{\rm m}32^{\rm s}.16$ & $-24^{\circ}35^{\prime}03.\!^{\prime\prime}2$ & $0.\!^{\prime\prime}76$ & $29^{+6}_{-5}$ & $0.17\pm0.24$ \\
4 & 2.96 & $17^{\rm h}17^{\rm m}07^{\rm s}.74$ & $-24^{\circ}34^{\prime}49.\!^{\prime\prime}3$ & $1.\!^{\prime\prime}41$ & $2.6^{+2.9}_{-1.7}$ & $<0.67$ \\
5 & 2.97 & $17^{\rm h}17^{\rm m}12^{\rm s}.49$ & $-24^{\circ}33^{\prime}26.\!^{\prime\prime}6$ & $1.\!^{\prime\prime}41$ & $2.6^{+2.9}_{-1.7}$ & $<0.67$ \\
6 & 4.18 & $17^{\rm h}17^{\rm m}23^{\rm s}.94$ & $-24^{\circ}31^{\prime}49.\!^{\prime\prime}3$ & $1.\!^{\prime\prime}43$ & $5.2^{+3.6}_{-2.4}$ & $<0.13$ \\
7 & 4.95 & $17^{\rm h}16^{\rm m}58^{\rm s}.26$ & $-24^{\circ}36^{\prime}29.\!^{\prime\prime}2$ & $0.\!^{\prime\prime}90$ & $31^{+7}_{-6}$ & $0.11\pm0.24$ \\
8 & 5.24 & $17^{\rm h}17^{\rm m}42^{\rm s}.86$ & $-24^{\circ}36^{\prime}12.\!^{\prime\prime}6$ & $1.\!^{\prime\prime}34$ & $11^{+5}_{-3}$ & $0.30\pm0.49$ \\
9 & 5.41 & $17^{\rm h}17^{\rm m}39^{\rm s}.59$ & $-24^{\circ}38^{\prime}55.\!^{\prime\prime}5$ & $2.\!^{\prime\prime}40$ & $4.6^{+3.6}_{-2.4}$ & $>-0.12$ \\
10 & 6.95 & $17^{\rm h}17^{\rm m}24^{\rm s}.06$ & $-24^{\circ}42^{\prime}46.\!^{\prime\prime}8$ & $2.\!^{\prime\prime}79$ & $7.7^{+4.6}_{-3.4}$ & $>-0.07$ \\
11 & 7.01 & $17^{\rm h}17^{\rm m}40^{\rm s}.13$ & $-24^{\circ}30^{\prime}36.\!^{\prime\prime}2$ & $2.\!^{\prime\prime}85$ & $7.7^{+4.6}_{-3.4}$ & $-0.08\pm0.66$ \\
12 & 7.05 & $17^{\rm h}17^{\rm m}02^{\rm s}.98$ & $-24^{\circ}29^{\prime}58.\!^{\prime\prime}7$ & $1.\!^{\prime\prime}70$ & $17^{+6}_{-5}$ & $0.14\pm0.37$ \\
13 & 7.12 & $17^{\rm h}16^{\rm m}48^{\rm s}.88$ & $-24^{\circ}34^{\prime}51.\!^{\prime\prime}1$ & $2.\!^{\prime\prime}49$ & $10^{+5}_{-4}$ & $0.34\pm0.58$ \\
14 & 7.99 & $17^{\rm h}17^{\rm m}13^{\rm s}.11$ & $-24^{\circ}43^{\prime}44.\!^{\prime\prime}3$ & $4.\!^{\prime\prime}11$ & $7.4^{+4.7}_{-3.6}$ & $<0.41$ \\ \hline
\multicolumn{7}{c}{IGR J17306--2015}\\ \hline
1 & 1.26 & $17^{\rm h}30^{\rm m}24^{\rm s}.81$ & $-20^{\circ}17^{\prime}03.\!^{\prime\prime}3$ & $0.\!^{\prime\prime}75$ & $12^{+5}_{-3}$ & $0.16\pm0.43$ \\
2 & 1.45 & $17^{\rm h}30^{\rm m}20^{\rm s}.40$ & $-20^{\circ}14^{\prime}36.\!^{\prime\prime}3$ & $0.\!^{\prime\prime}88$ & $3.9^{+3.2}_{-1.9}$ & $>-0.25$ \\
3 & 2.91 & $17^{\rm h}30^{\rm m}30^{\rm s}.27$ & $-20^{\circ}13^{\prime}20.\!^{\prime\prime}0$ & $0.\!^{\prime\prime}95$ & $6.8^{+3.8}_{-2.6}$ & $-0.16\pm0.65$ \\
4 & 3.04 & $17^{\rm h}30^{\rm m}33^{\rm s}.99$ & $-20^{\circ}13^{\prime}55.\!^{\prime\prime}8$ & $1.\!^{\prime\prime}42$ & $2.8^{+2.9}_{-1.7}$ & --- \\
5 & 6.66 & $17^{\rm h}29^{\rm m}58^{\rm s}.31$ & $-20^{\circ}12^{\prime}53.\!^{\prime\prime}0$ & $2.\!^{\prime\prime}68$ & $7.2^{+4.3}_{-3.1}$ & $<-0.13$ \\
6 & 6.76 & $17^{\rm h}30^{\rm m}20^{\rm s}.66$ & $-20^{\circ}22^{\prime}32.\!^{\prime\prime}2$ & $2.\!^{\prime\prime}78$ & $7.1^{+4.3}_{-3.1}$ & $<-0.44$ \\
7 & 7.96 & $17^{\rm h}30^{\rm m}50^{\rm s}.39$ & $-20^{\circ}10^{\prime}52.\!^{\prime\prime}1$ & $2.\!^{\prime\prime}73$ & $12^{+5}_{-4}$ & $-0.01\pm0.47$ \\ \hline
\multicolumn{7}{c}{IGR J18088--2741}\\ \hline
1 & 0.38 & $18^{\rm h}08^{\rm m}58^{\rm s}.55$ & $-27^{\circ}41^{\prime}48.\!^{\prime\prime}9$ & $0.\!^{\prime\prime}76$ & $6.9^{+3.8}_{-2.6}$ & $-0.15\pm0.64$ \\
2 & 1.42 & $18^{\rm h}09^{\rm m}06^{\rm s}.57$ & $-27^{\circ}42^{\prime}00.\!^{\prime\prime}2$ & $0.\!^{\prime\prime}88$ & $3.8^{+3.2}_{-1.9}$ & $<0.96$ \\
3 & 1.65 & $18^{\rm h}08^{\rm m}53^{\rm s}.76$ & $-27^{\circ}42^{\prime}34.\!^{\prime\prime}8$ & $0.\!^{\prime\prime}86$ & $4.8^{+3.4}_{-2.2}$ & $<0.29$ \\
4 & 1.68 & $18^{\rm h}09^{\rm m}03^{\rm s}.59$ & $-27^{\circ}40^{\prime}15.\!^{\prime\prime}6$ & $0.\!^{\prime\prime}83$ & $5.8^{+3.6}_{-2.4}$ & $>-0.41$ \\
5 & 2.21 & $18^{\rm h}08^{\rm m}50^{\rm s}.40$ & $-27^{\circ}41^{\prime}26.\!^{\prime\prime}1$ & $0.\!^{\prime\prime}84$ & $7.8^{+4.0}_{-2.8}$ & $<0.11$ \\
6 & 2.21 & $18^{\rm h}08^{\rm m}58^{\rm s}.14$ & $-27^{\circ}39^{\prime}36.\!^{\prime\prime}6$ & $0.\!^{\prime\prime}77$ & $16^{+5}_{-4}$ & $-0.01\pm0.36$ \\
7 & 2.27 & $18^{\rm h}09^{\rm m}07^{\rm s}.57$ & $-27^{\circ}43^{\prime}21.\!^{\prime\prime}5$ & $0.\!^{\prime\prime}94$ & $4.8^{+3.4}_{-2.2}$ & $<0.30$ \\
8 & 2.29 & $18^{\rm h}08^{\rm m}58^{\rm s}.24$ & $-27^{\circ}44^{\prime}00.\!^{\prime\prime}9$ & $0.\!^{\prime\prime}94$ & $4.8^{+3.4}_{-2.2}$ & $<0.30$ \\
9 & 2.61 & $18^{\rm h}08^{\rm m}50^{\rm s}.82$ & $-27^{\circ}40^{\prime}12.\!^{\prime\prime}6$ & $1.\!^{\prime\prime}00$ & $4.7^{+3.4}_{-2.2}$ & $<0.30$ \\
10 & 2.79 & $18^{\rm h}08^{\rm m}48^{\rm s}.86$ & $-27^{\circ}40^{\prime}34.\!^{\prime\prime}6$ & $0.\!^{\prime\prime}97$ & $5.7^{+3.6}_{-2.4}$ & $>-0.43$ \\
11 & 3.25 & $18^{\rm h}09^{\rm m}14^{\rm s}.60$ & $-27^{\circ}41^{\prime}03.\!^{\prime\prime}5$ & $0.\!^{\prime\prime}79$ & $24^{+6}_{-5}$ & $0.16\pm0.28$ \\
12 & 3.50 & $18^{\rm h}09^{\rm m}10^{\rm s}.00$ & $-27^{\circ}44^{\prime}19.\!^{\prime\prime}9$ & $1.\!^{\prime\prime}15$ & $5.5^{+3.6}_{-2.4}$ & $<-0.08$ \\
13 & 3.52 & $18^{\rm h}08^{\rm m}47^{\rm s}.45$ & $-27^{\circ}43^{\prime}51.\!^{\prime\prime}1$ & $1.\!^{\prime\prime}03$ & $7.5^{+4.0}_{-2.8}$ & $<0.11$ \\
14 & 3.54 & $18^{\rm h}08^{\rm m}49^{\rm s}.93$ & $-27^{\circ}44^{\prime}28.\!^{\prime\prime}1$ & $1.\!^{\prime\prime}28$ & $4.5^{+3.4}_{-2.2}$ & $<0.33$ \\
15 & 3.61 & $18^{\rm h}09^{\rm m}14^{\rm s}.30$ & $-27^{\circ}43^{\prime}36.\!^{\prime\prime}4$ & $1.\!^{\prime\prime}50$ & $3.5^{+3.1}_{-1.9}$ & $<0.33$ \\
16 & 4.02 & $18^{\rm h}08^{\rm m}53^{\rm s}.34$ & $-27^{\circ}38^{\prime}03.\!^{\prime\prime}3$ & $1.\!^{\prime\prime}25$ & $6.2^{+3.8}_{-2.6}$ & $-0.20\pm0.71$ \\
17 & 4.05 & $18^{\rm h}08^{\rm m}57^{\rm s}.42$ & $-27^{\circ}37^{\prime}45.\!^{\prime\prime}8$ & $1.\!^{\prime\prime}18$ & $7.2^{+4.0}_{-2.8}$ & $0.24\pm0.64$ \\
18 & 4.24 & $18^{\rm h}08^{\rm m}45^{\rm s}.39$ & $-27^{\circ}44^{\prime}26.\!^{\prime\prime}2$ & $0.\!^{\prime\prime}87$ & $23^{+6}_{-5}$ & $-0.36\pm0.29$ \\
19 & 4.53 & $18^{\rm h}08^{\rm m}39^{\rm s}.80$ & $-27^{\circ}41^{\prime}31.\!^{\prime\prime}6$ & $0.\!^{\prime\prime}74$ & $230^{+16}_{-15}$ & $0.76\pm0.09$ \\
20 & 4.67 & $18^{\rm h}09^{\rm m}20^{\rm s}.65$ & $-27^{\circ}42^{\prime}57.\!^{\prime\prime}5$ & $1.\!^{\prime\prime}56$ & $5.9^{+3.8}_{-2.6}$ & $-0.24\pm0.75$ \\
21 & 6.24 & $18^{\rm h}09^{\rm m}28^{\rm s}.33$ & $-27^{\circ}41^{\prime}11.\!^{\prime\prime}9$ & $2.\!^{\prime\prime}71$ & $5.9^{+4.0}_{-2.8}$ & $<-0.07$ \\
22 & 6.25 & $18^{\rm h}09^{\rm m}00^{\rm s}.14$ & $-27^{\circ}48^{\prime}00.\!^{\prime\prime}9$ & $1.\!^{\prime\prime}75$ & $11^{+5}_{-4}$ & $-0.16\pm0.49$ \\
23 & 6.30 & $18^{\rm h}08^{\rm m}56^{\rm s}.93$ & $-27^{\circ}48^{\prime}01.\!^{\prime\prime}5$ & $2.\!^{\prime\prime}77$ & $5.9^{+4.0}_{-2.8}$ & $-0.13\pm0.77$ \\
24 & 6.76 & $18^{\rm h}08^{\rm m}30^{\rm s}.22$ & $-27^{\circ}40^{\prime}33.\!^{\prime\prime}0$ & $2.\!^{\prime\prime}40$ & $8.6^{+4.6}_{-3.5}$ & $>-0.10$ \\
25 & 7.40 & $18^{\rm h}09^{\rm m}00^{\rm s}.16$ & $-27^{\circ}49^{\prime}10.\!^{\prime\prime}3$ & $1.\!^{\prime\prime}37$ & $30^{+7}_{-6}$ & $-0.29\pm0.26$ \\
26 & 7.60 & $18^{\rm h}08^{\rm m}36^{\rm s}.70$ & $-27^{\circ}47^{\prime}17.\!^{\prime\prime}8$ & $3.\!^{\prime\prime}71$ & $7.2^{+4.6}_{-3.5}$ & $>-0.21$ \\
27 & 7.72 & $18^{\rm h}08^{\rm m}30^{\rm s}.25$ & $-27^{\circ}45^{\prime}42.\!^{\prime\prime}7$ & $2.\!^{\prime\prime}39$ & $13^{+5}_{-4}$ & $0.17\pm0.45$ \\
28 & 7.87 & $18^{\rm h}08^{\rm m}54^{\rm s}.52$ & $-27^{\circ}49^{\prime}32.\!^{\prime\prime}1$ & $1.\!^{\prime\prime}91$ & $20^{+6}_{-5}$ & $-0.25\pm0.35$ \\
29 & 9.96 & $18^{\rm h}08^{\rm m}38^{\rm s}.95$ & $-27^{\circ}50^{\prime}32.\!^{\prime\prime}5$ & $12.\!^{\prime\prime}77$ & $4.5^{+5.5}_{-4.4}$ & $<0.74$ \\ \hline
\multicolumn{7}{c}{IGR J18381--0924}\\ \hline
1 & 1.84 & $18^{\rm h}38^{\rm m}02^{\rm s}.02$ & $-09^{\circ}23^{\prime}43.\!^{\prime\prime}9$ & $0.\!^{\prime\prime}73$ & $33^{+7}_{-6}$ & $0.70\pm0.26$ \\
2 & 2.78 & $18^{\rm h}38^{\rm m}18^{\rm s}.58$ & $-09^{\circ}25^{\prime}52.\!^{\prime\prime}2$ & $0.\!^{\prime\prime}70$ & $303^{+18}_{-17}$ & $0.73\pm0.08$ \\
3 & 2.82 & $18^{\rm h}38^{\rm m}20^{\rm s}.11$ & $-09^{\circ}23^{\prime}55.\!^{\prime\prime}6$ & $1.\!^{\prime\prime}14$ & $3.7^{+3.2}_{-1.9}$ & $<0.54$ \\
4 & 3.63 & $18^{\rm h}38^{\rm m}15^{\rm s}.91$ & $-09^{\circ}27^{\prime}38.\!^{\prime\prime}9$ & $1.\!^{\prime\prime}30$ & $4.6^{+3.4}_{-2.2}$ & $<0.61$ \\
5 & 3.76 & $18^{\rm h}38^{\rm m}16^{\rm s}.55$ & $-09^{\circ}21^{\prime}12.\!^{\prime\prime}6$ & $1.\!^{\prime\prime}14$ & $6.5^{+3.8}_{-2.6}$ & $<0.00$ \\
6 & 4.14 & $18^{\rm h}38^{\rm m}22^{\rm s}.64$ & $-09^{\circ}22^{\prime}05.\!^{\prime\prime}8$ & $0.\!^{\prime\prime}80$ & $42^{+8}_{-7}$ & $>0.58$ \\
7 & 4.27 & $18^{\rm h}37^{\rm m}51^{\rm s}.59$ & $-09^{\circ}24^{\prime}09.\!^{\prime\prime}6$ & $1.\!^{\prime\prime}24$ & $7.4^{+4.0}_{-2.8}$ & $<-0.10$ \\
8 & 5.72 & $18^{\rm h}37^{\rm m}59^{\rm s}.66$ & $-09^{\circ}29^{\prime}42.\!^{\prime\prime}5$ & $1.\!^{\prime\prime}76$ & $8.4^{+4.3}_{-3.1}$ & $>0.13$ \\
9 & 6.14 & $18^{\rm h}38^{\rm m}30^{\rm s}.63$ & $-09^{\circ}27^{\prime}26.\!^{\prime\prime}9$ & $1.\!^{\prime\prime}77$ & $10^{+5}_{-3}$ & $<-0.11$ \\
10 & 6.46 & $18^{\rm h}38^{\rm m}07^{\rm s}.59$ & $-09^{\circ}18^{\prime}00.\!^{\prime\prime}3$ & $3.\!^{\prime\prime}47$ & $4.8^{+3.8}_{-2.6}$ & $<0.25$ \\
11 & 6.49 & $18^{\rm h}38^{\rm m}33^{\rm s}.67$ & $-09^{\circ}26^{\prime}38.\!^{\prime\prime}1$ & $2.\!^{\prime\prime}02$ & $10^{+5}_{-3}$ & $-0.13\pm0.52$ \\
12 & 7.51 & $18^{\rm h}38^{\rm m}17^{\rm s}.63$ & $-09^{\circ}31^{\prime}39.\!^{\prime\prime}2$ & $2.\!^{\prime\prime}76$ & $10^{+5}_{-4}$ & $-0.05\pm0.54$ \\
13 & 8.16 & $18^{\rm h}37^{\rm m}50^{\rm s}.30$ & $-09^{\circ}17^{\prime}41.\!^{\prime\prime}9$ & $1.\!^{\prime\prime}11$ & $81^{+10}_{-9}$ & $0.01\pm0.13$ \\
14 & 8.37 & $18^{\rm h}37^{\rm m}54^{\rm s}.41$ & $-09^{\circ}32^{\prime}01.\!^{\prime\prime}7$ & $3.\!^{\prime\prime}24$ & $11^{+5}_{-4}$ & $-0.25\pm0.51$ \\ \hline
\multicolumn{7}{c}{IGR J20107+4534}\\ \hline
1 & 1.40 & $20^{\rm h}10^{\rm m}43^{\rm s}.20$ & $+45^{\circ}34^{\prime}49.\!^{\prime\prime}6$ & $0.\!^{\prime\prime}76$ & $11^{+4}_{-3}$ & $-0.28\pm0.47$ \\
2 & 1.44 & $20^{\rm h}10^{\rm m}30^{\rm s}.05$ & $+45^{\circ}34^{\prime}46.\!^{\prime\prime}7$ & $0.\!^{\prime\prime}80$ & $6.8^{+3.8}_{-2.6}$ & $<-0.19$ \\
3 & 2.12 & $20^{\rm h}10^{\rm m}48^{\rm s}.95$ & $+45^{\circ}34^{\prime}06.\!^{\prime\prime}5$ & $0.\!^{\prime\prime}74$ & $26^{+6}_{-5}$ & $<-0.66$ \\
4 & 2.87 & $20^{\rm h}10^{\rm m}20^{\rm s}.61$ & $+45^{\circ}33^{\prime}34.\!^{\prime\prime}7$ & $1.\!^{\prime\prime}05$ & $4.7^{+3.4}_{-2.2}$ & $>-0.66$ \\
5 & 3.63 & $20^{\rm h}10^{\rm m}47^{\rm s}.47$ & $+45^{\circ}30^{\prime}51.\!^{\prime\prime}2$ & $1.\!^{\prime\prime}31$ & $4.5^{+3.4}_{-2.2}$ & $>-0.11$ \\
6 & 3.69 & $20^{\rm h}10^{\rm m}34^{\rm s}.82$ & $+45^{\circ}37^{\prime}38.\!^{\prime\prime}3$ & $1.\!^{\prime\prime}90$ & $2.5^{+2.9}_{-1.7}$ & $<0.77$ \\
7 & 4.14 & $20^{\rm h}10^{\rm m}25^{\rm s}.50$ & $+45^{\circ}37^{\prime}36.\!^{\prime\prime}1$ & $1.\!^{\prime\prime}40$ & $5.3^{+3.6}_{-2.4}$ & $>-0.49$ \\
8 & 4.30 & $20^{\rm h}10^{\rm m}12^{\rm s}.27$ & $+45^{\circ}33^{\prime}55.\!^{\prime\prime}6$ & $1.\!^{\prime\prime}35$ & $6.2^{+3.8}_{-2.6}$ & $-0.22\pm0.71$ \\
9 & 4.54 & $20^{\rm h}10^{\rm m}52^{\rm s}.12$ & $+45^{\circ}30^{\prime}17.\!^{\prime\prime}8$ & $0.\!^{\prime\prime}99$ & $16^{+5}_{-4}$ & $<-0.49$ \\
10 & 4.74 & $20^{\rm h}10^{\rm m}26^{\rm s}.36$ & $+45^{\circ}38^{\prime}20.\!^{\prime\prime}4$ & $3.\!^{\prime\prime}36$ & $2.1^{+2.9}_{-1.7}$ & --- \\
11 & 5.20 & $20^{\rm h}10^{\rm m}56^{\rm s}.40$ & $+45^{\circ}37^{\prime}53.\!^{\prime\prime}7$ & $1.\!^{\prime\prime}47$ & $8.7^{+4.3}_{-3.1}$ & $-0.06\pm0.55$ \\
12 & 5.63 & $20^{\rm h}10^{\rm m}27^{\rm s}.57$ & $+45^{\circ}28^{\prime}34.\!^{\prime\prime}4$ & $1.\!^{\prime\prime}23$ & $16^{+5}_{-4}$ & $<-0.51$ \\
13 & 7.48 & $20^{\rm h}11^{\rm m}19^{\rm s}.30$ & $+45^{\circ}34^{\prime}51.\!^{\prime\prime}8$ & $1.\!^{\prime\prime}38$ & $31^{+7}_{-6}$ & $0.10\pm0.25$ \\ \hline
\end{longtable}
\begin{minipage}[b]{\textwidth}
$^{a}$The angular distance between the center of the {\em INTEGRAL} error circle and the source.  In each case, the {\em INTEGRAL} error circle radius is $2.\!^{\prime}1$ at 1-$\sigma$ \citep{krivonos12} and $4.\!^{\prime}2$ at 2-$\sigma$.\\ \\ $^{b}$The 90\% confidence uncertainty on the position, including statistical and systematic contributions.\\ \\ $^{c}$The number of ACIS-I counts detected (after background subtraction) in the 0.3--10 keV band.  The errors are 68\% confidence Poisson errors using the analytical approximations from \cite{gehrels86}.\\ \\ $^{d}$The hardness is given by $(C_{2}-C_{1})/(C_{2}+C_{1})$, where $C_{2}$ is the number of counts in the 2--10 keV band and $C_{1}$ is the number of counts in the 0.3--2 keV band.
\end{minipage}
\end{center}
}

\begin{table}
\caption{Position Shifts Based on {\em Chandra}/near-IR Matches\label{tab:matches_and_shifts}}
\begin{minipage}{\linewidth}
\begin{center}
\footnotesize
\begin{tabular}{ccccc} \hline \hline
IGR Name & Near-IR Catalog & $N_{\rm matches}$\footnote{The number of matches between {\em Chandra} sources and sources in the near-IR catalog after running {\ttfamily wcs\_match}.} & R.A. shift & Decl. shift\\ \hline\hline
J14091--6108 & VISTA/VVV & 10 &  +$0.\!^{\prime\prime}01$ &  +$0.\!^{\prime\prime}23$\\
J15335--5420 & VISTA/VVV & 5  & --$0.\!^{\prime\prime}01$ & --$0.\!^{\prime\prime}25$\\
J17164--3803 & VISTA/VVV & 5  & --$0.\!^{\prime\prime}43$ & --$0.\!^{\prime\prime}20$\\
J17174--2436 & UKIDSS    & 1  &   ---                    & ---\\
J17306--2015 & 2MASS     & 2  &   ---                    & ---\\
J18088--2741 & VISTA/VVV & 15 & --$0.\!^{\prime\prime}09$ &  +$0.\!^{\prime\prime}09$\\
J18381--0924 & UKIDSS    & 3  & ---                      & ---\\
J20107+4534  & 2MASS     & 5  & --$0.\!^{\prime\prime}29$ & --$0.\!^{\prime\prime}29$\\ \hline
\end{tabular}
\end{center}
\end{minipage}
\end{table}

\end{document}